\documentclass[final, 12pt, a4paper, english]{article}

\usepackage[
  tracking=true,
  expansion=true,
  protrusion=true,
  babel,
  nopatch=footnote
]{microtype}

\usepackage{ifthen}
\newif\ifdelete
\deletetrue     % use \deletetrue to remove parts, use \deletefalse to have these parts in red

\usepackage{environ}
\NewEnviron{delete}{
    \ifdelete
    \else
    \bigskip
    \color{red}
    \par\noindent\rule{\textwidth}{0.4pt}
    \centerline{begin delete}
    \BODY
    \par\noindent\rule{\textwidth}{0.4pt}
    \centerline{end delete}
    \color{black}
    \fi
}{%
}%

\usepackage[utf8]{inputenc}
\usepackage[T1]{fontenc}
\usepackage[english]{babel}
\usepackage[
  a4paper,
  top=2.75cm,
  left=2.75cm,
  right=2.75cm,
  bottom=3cm,
]{geometry}

\usepackage{titletoc}

\renewcommand{\tableofcontents}{
  \startcontents[main]
  \printcontents[main]{l}{1}{}
}
\newcommand{\restarttoc}{
  \stopcontents[main]
  \renewcommand{\contentsname}{Supplement Contents}
  \clearpage
  \startcontents[supplement]
}
\newcommand{\supplementtoc}{
  \printcontents[supplement]{l}{1}{}
}

\usepackage{supertabular}
\usepackage{multirow}
\usepackage{sidecap}
\sidecaptionvpos{figure}{t}
\usepackage{amsmath}
\usepackage{physics}
\usepackage{amsfonts}
\usepackage{amsthm}
\usepackage{amssymb}
\usepackage{textcomp}
\usepackage{gensymb}
\usepackage[table,x11names]{xcolor}
\usepackage{lmodern}
\usepackage{subfig}
\usepackage{fix-cm}
\usepackage[list=no]{caption}
\usepackage{graphicx}
\usepackage{booktabs}
\usepackage[mathlines]{lineno}
\usepackage{changepage}
\usepackage{floatrow}
\usepackage{helvet}
\usepackage{xfrac}
\usepackage{mathpazo}

\usepackage{sfmath}

\usepackage{enumitem}

\usepackage{nth}
\usepackage{tikz}
\usetikzlibrary{arrows, arrows.meta, calc, positioning, quotes, shapes, decorations.pathreplacing}
\usepackage{comment}
\usepackage{placeins}
\usepackage{orcidlink}

\usepackage{hyperref}
\usepackage[nameinlink]{cleveref}
\crefformat{figure}{#2Fig.~#1#3}

\crefformat{supfigure}{#2Supplementary Fig.~#1#3}
\Crefformat{supfigure}{#2Supplementary Figure~#1#3}
\crefmultiformat{supfigure}{Supplementary Figs.~#2#1#3}{ and~#2#1#3}{, #2#1#3}{ and~#2#1#3}
\crefrangeformat{supfigure}{Supplementary Figs. #3#1#4--#5#2#6}{}{}{}
\crefformat{suptable}{#2Supplementary Tbl.~#1#3}
\Crefformat{suptable}{#2Supplementary Table~#1#3}

\graphicspath{{./graphics/}}
\DeclareRobustCommand{\ie}{i.\,e.~}

\newif\ifusesubsubsubsections
\usesubsubsubsectionsfalse

\hypersetup{colorlinks=true}
\hypersetup{linkcolor=red!50!black}
\hypersetup{urlcolor=blue!50!black}
\hypersetup{citecolor=green!50!black}

\hypersetup{pdfsubject={}}
\hypersetup{pdfkeywords={}{}{}}

\usepackage{setspace}
\newcommand{\articletitle}{Intergenerational Equitable Climate Change Mitigation:\newline Negative Effects of Stochastic Interest Rates and;\newline Positive Effects of Financing}

\hypersetup{pdftitle={Intergenerational Equitable Climate Change Mitigation: Negative Effects of Stochastic Interest Rates; Positive Effects of Financing}}

\begin{document}

%TC:ignore

  {
    \large
    \noindent
    \articletitle
  }
  \vspace{1em}

  {
    \noindent \small
    Christian P.~Fries \textsuperscript{a,b * $\dagger$} \orcidlink{0000-0003-4767-2034}
    \,
    and
    \,
    Lennart Quante \textsuperscript{c,d *} \orcidlink{0000-0003-4942-8254}
    
  }
  \vspace{0.5em}

  {
    \footnotesize
    \noindent
    \begin{enumerate}[topsep=0pt,itemsep=0ex,partopsep=0ex,parsep=0ex,leftmargin=*]
        \item[\textsuperscript{a}] Department of Mathematics, Ludwig Maximilians University, Munich, Germany

        \item[\textsuperscript{b}] DZ Bank AG Deutsche Zentral-Genossenschaftsbank, Frankfurt a. M., Germany

        \item[\textsuperscript{c}] Potsdam Institute for Climate Impact Research (PIK), Member of the Leibniz\\ Association, Potsdam, Germany

        \item[\textsuperscript{d}] Institute of Mathematics, University of Potsdam, Potsdam, Germany

        \item[*] These authors contributed equally to this work.

        \item[$\dagger$] Correspondence to: email@christian-fries.de
    \end{enumerate}
}

% The following block is only for online version (not for submission)
\vspace{0.5em}
{
    \footnotesize
	\noindent
    December 12th, 2023 [v1]. March 28th, 2024 [v2], December 23rd, 2024 [v3].\newline
    This version: August 12th, 2025 [v4]
}

\thispagestyle{empty}

\vspace{1.5em}

%\begin{abstract}
    
\section*{Abstract}
{
\setlength{\parskip}{1.75pt}\sloppy
Climate mitigation decisions today affect future generations, raising questions of intergenerational equity.
Integrated assessment models (IAMs) rely on discounting to evaluate long-term policy costs and benefits. Using the DICE model, we quantify how optimal pathways distribute abatement and damage costs across cohorts. Unconstrained optimization creates intergenerational inequality, with future generations bearing higher costs relative to GDP. Extending the model with stochastic discount rates, we show that discount-rate uncertainty significantly amplifies this inequality.
We consider two independent extensions: the financing of abatement costs and the modeling of nonlinear financing costs under large damages. Both extensions can materially improve intergenerational equity by distributing mitigation efforts more evenly. As an illustration, we present a modified DICE model whose optimal pathway limits generational costs to 3~\% of GDP, leading to more equitable effort sharing.
Our proposed model extensions are model-agnostic, applicable across IAMs, and compatible with alternative intergenerational equity metrics. 
}
%\end{abstract}

%
% Old abstract
%

\begin{comment}
    
Today's decisions on climate change mitigation will affect the damage that future generations will bear. Discounting the future benefits and costs of climate change mitigation is one of the most critical components to evaluate efficient climate mitigation pathways. 
We examine cost-based intergenerational equity and its interaction with interest rate uncertainty in integrated assessment models (IAMs), using DICE as a commonly used illustrative example.
Stochastic rates give rise to a stochastic mitigation strategy, resulting in all model quantities becoming stochastic.

We show that the optimization procedure of the DICE model induces intergenerational inequality: lacking a mechanism to regulate burden, future generations must bear higher costs from abatement and damage relative to GDP.

Further, we show that considering the uncertainty of discount rates and their feedback on abatement policies, which can be interpreted as successive re-calculation, increases intergenerational inequality (and adds additional risks).
Motivated by this, we consider additional financing risks by investigating two modifications of DICE. We find that allowing financing of abatement costs and considering non-linear financing effects for large damages improves intergenerational effort sharing. To conclude our discussion of options to improve intergenerational equity in an IAM, we propose a modified optimization to keep costs below 3\% of GDP, resulting in more equal distribution of efforts between generations.

\end{comment}

\newpage
\tableofcontents

%TC:endignore

\newpage

\section{Introduction}
Climate change is one of the largest risks of the next centuries \cite{masson-delmotte_climate_2021}, as demonstrated by its manifold impacts on society \cite{portner_ipcc_2022}. Owing to the persistent nature of the main greenhouse gas carbon dioxide \cite{solomon_persistence_2010} on human time scales, a transition to (net) zero emissions is necessary to limit global warming to any constant level. A well-established method for the assessment of transition pathways are integrated assessment models (IAMs) of climate and economic systems \cite{weyant_contributions_2017,nordhaus_evolution_2018}. 

We focus on the DICE model \cite{Nordhaus2013, Nordhaus2017,barrage_policies_2024} as a simple IAM, which is frequently used to demonstrate the impacts of modifications to the original model due to its simplicity \cite{faulwasser_towards_2018,wirths_permafrost_2018,Grubb2021ModelingMO,Glanemann2020,Hansel2020}. 
While IAMs were first set up as deterministic, stochastic shocks have been included to consider the risk of tipping points \cite{crost_optimal_2013,lemoine_watch_2014,jensen_optimal_2014,cai_environmental_2015,cai_risk_2016,lemoine_economics_2016,dietz_economic_2021}, natural feedback processes such as permafrost \cite{wirths_permafrost_2018} or abstract catastrophic risks \cite{Ackerman2010a, ikefuji_expected_2020}.
Because most of the damages from climate change will occur over long time scales, value-based decisions about the discount rate and the optimization function have a strong influence on the resulting optimal paths \cite{Hepburn2007, Gollier2010, van_der_ploeg_simple_2019,VanDerPloeg2020,wagner_carbon_2020}. Stochastic approaches also require more advanced risk evaluations than Monte Carlo averaging to capture the full extent of tail risks \cite{crost_optimal_2013} or inequalities in the distribution of damages \cite{adler_priority_2017}.

% SCC Section of Intro moved to Supplement

Discount rates, in particular, are an essential component of IAM modeling. Thus, the discussion between proponents of a descriptive approach based on observation of market returns as in DICE \cite{Nordhaus2017} versus a normative approach to discounting as in the Stern review \cite{Hepburn2007} continues. The assumptions on discounting have already been discussed in the context of intergenerational \cite{padilla_intergenerational_2002} and intragenerational inequality \cite{emmerling_discounting_2018}. Recent studies \cite{bauer_rising_2021,newell_discounting_2022} show that dynamic modeling of discount rates might allow a more precise assessment of the costs of damage and abatement. While our focus lies on the effect of uncertainty about discount rates and stochastic adaptation of abatement policies on intergenerational equity, we acknowledge that intergenerational equity can be tackled with other policy extensions \cite{ANDERSEN2020102290}, e.g. carbon taxation, \cite{Kotlikoff2021CarbonTaxation}.

%note on DICE 2023 -> specific introduction
Parallel to this work, an updated DICE-2023 model was developed and published \cite{barrage_policies_2024} - one of the most prominent updates includes the consideration of uncertainty in the discount rate through the inclusion of consumption growth uncertainty and riskiness of climate investments. These uncertainties are translated into certainty-equivalent discount rate, such that the overall model remains deterministic. 
As we show in \cref{eq:dice:stochastic_rates_in_deterministic_model}, even a stochastic interest rate without feedback processes within the model is equivalent to a deterministic approach.

In contrast to this deterministic treatment of interest rate risk, we consider uncertainty about the rate of time preference $\rho$ by modeling it as a stochastic process using a standard interest rate model. Introducing adaptation of the abatement policy, all quantities become stochastic.

\medskip

We then show that the optimal mitigation pathway derived from an IAM may exhibit substantial intergenerational cost imbalances, which are critically exacerbated by uncertainties in interest rates. We propose two novel model extensions that markedly improve cost-based intergenerational equity, thus enhancing the robustness and fairness of modeling global climate mitigation, providing a crucial political impulse for effective climate action. \cref{fig:scheme_extensions} provides an overview of the extensions.

Here, we interpret intergenerational equity as ensuring that no cohort bears a disproportionate share of climate-related mitigation and damage costs relative to its economic capacity. To allow meaningful comparisons between generations of differing wealth and productivity, we measure costs as a percentage of GDP, following principles in public finance for sustainable debt and investment burdens \cite{daniel_pushing_2013}. As an illustrative benchmark rather than a normative prescription, we adopt a limit of 3 \% of GDP for combined mitigation and damage costs, drawing on similar proposals for a ``fair share'' of national income devoted to climate action and its consistency with typical upper bounds for long-term public investment commitments. We recognise that alternative justice principles, such as egalitarian, prioritarian, or sufficiency-based approaches \cite{padilla_intergenerational_2002}, would lead to different equity metrics and thresholds. Our results are therefore one contribution to the broader debate on embedding intergenerational fairness in integrated assessment models, but the model extensions we propose can be readily combined with such alternative metrics.

%TC:ignore
\begin{delete}

\section{Method: Extensions to the DICE Model}

%paragraph on methods (300 words)
We examine the optimization of the DICE model with respect to uncertainty in the rate of time-preference, and its consequences for intergenerational equity. Measuring the costs per GDP (\cref{sec:costPerGDP}) of the optimal (calibrated) climate emissions path in the standard model, we find that these costs are unequally distributed across generations.
We use an implementation of the DICE model with a flexible time discretization \cite{fries_intergenerational_2023} and couple this model with a comprehensive modeling toolbox for financial markets \cite{finmathlib} to enable the integration of valuation methods for complex financial products in IAMs.

\subsection{Interest Rates}

We add stochastic interest rates - through modeling the rate of time preference $\rho$ as stochastic - to the model, and introduce a stochastic abatement policy. This policy mimics the process of reacting to changes in the interest rate level by repeated (over time) re-calculation of the model.
Since intergenerational inequality is related to the sensitivity of the utility and the discount factor, we introduce two model extensions related to these two parts that influence intergenerational equity: Financing of abatement costs and non-linear discounting of costs, \cite{fries_non-linear_2021}. As abatement is a planned investment process, we assume that these costs may be financed by loans with uncertain interest rates. Furthermore, as large projects and investments often incur additional costs \cite{hallegatte_why_2007,shane_construction_2009}, we introduce increasing financing costs for large payments via a non-linear discounting. These higher costs for high damages can be viewed as (additional) financing risks.
\end{delete}
%TC:endignore

\begin{figure}[htb]
    \centering
    % \documentclass[12pt]{article}

% \input{preamble}
% \input{definitions}

% \begin{document}
% \begin{figure}

\usetikzlibrary{fit}

\centering

\definecolor{teal}{HTML}{3c93c2}
\definecolor{green}{HTML}{6CBC90}
\definecolor{grey}{HTML}{a0a0a0}
\definecolor{orange}{HTML}{fd8d3c}
\definecolor{purple}{HTML}{4b3b55}

\newcommand{\nodeDistance}{0.2cm}

\newcommand{\wideWidth}{14.5cm}
\newcommand{\boxesWidth}{4.5cm}

\newcommand{\boxalign}{left}
\newcommand{\innersep}{3pt}

\tikzset{
  box/.style = {
    shape = rectangle,
    align = \boxalign,
    rounded corners,
    minimum width = 10em,
    minimum height = 2em,
    draw = black!40,
    line width = 1pt,
    font=\small,
    inner sep=\innersep,
  },
  doublebox/.style = {
    box,
    double,
    double distance = .5mm
  },
  diamond/.style = {
    shape = diamond,
    align = center,
    minimum width = 10em,
    minimum height = 5em,
    draw = black!40,
    line width = 1pt
  },
  diamond_grey/.style = {
    diamond,
    draw  = lightgrey,
    fill  = lightgrey!30
  },
  colored box/.style = {
    box,
    draw  = #1!,
    fill  = #1!35
  },
  arrow/.style = {
    {-Stealth[length=10pt, width=7pt]},
    shorten <= 0.5em,
    shorten >= 0.5em,
    line width=1.3pt,
    align=left,
    line width = 1pt,
    color=black!70!white,
  },
  darrow/.style = {
    arrow,
    <->
  },
  dotarrow/.style = {
    arrow,
    dotted
  },
  hv/.style = {to path = {-|(\tikztotarget)\tikztonodes}},
  vh/.style = {to path = {|-(\tikztotarget)\tikztonodes}},
  fat blue line/.style = {ultra thick, blue}
}

\begin{tikzpicture}[thick, >=latex, node distance=\nodeDistance]

\coordinate (start);

\node [text width = \wideWidth, colored box={grey}, left=of start, label={}] (A)
{
\textbf{Basic model DICE-2016-R \cite{Nordhaus2017} \ -- \ re-implementation with improvements\\}
\vspace{1em}
\begin{minipage}{\wideWidth}
\begin{itemize}[leftmargin=1em,parsep=0pt]
\begin{footnotesize}
    \item Integrating numerical methods \cite{finmathlib} (Monte-Carlo simulation, algorithmic differentiation)
    \item Improved representation of time-discretisation via an Euler-scheme.
\end{footnotesize} 
\end{itemize}
\end{minipage}
};

\node [text width = \boxesWidth, colored box={green}, above right=0.25cm and 0cm of A.north west,align=left, label={}] (B)
{
\textbf{Stochastic \\ interest rates \phantom{g}}
%% TODO If we write such details here, should we also add details to the other boxes?
%\vspace{1em}
%\begin{minipage}{\boxesWidth}
%\begin{itemize}[leftmargin=1em,parsep=0pt]
%\begin{footnotesize}
%    \item stochastic rates with long term mean $\mu$
%    \item fluctuation controlled by volatility $\sigma$
%\end{footnotesize} 
%\end{itemize}
%\end{minipage}
};

\node [text width = \boxesWidth,colored box={orange}, above=of A.north, right=of B.east, align=left, label={}] (C)
{
\textbf{Funding of \\ abatement cost \phantom{g}}
};

\node [text width = \boxesWidth,colored box={teal}, above=of A.north, right=of C.east,align=left,align=left, label={}] (D)
{
\textbf{Non-linear funding of \\ large damage cost}
};

\node [text width = \wideWidth,colored box={blue}, above right=0.25cm and 0cm of B.north west,align=left,align=left, label={}] (F)
{
\textbf{Limiting total cost by GDP share}
};

%\node [draw, inner sep=5pt, fit=(A)(F)] (boundingbox) {};
\clip[draw, inner sep=5pt, fit=(A)(F)];

\end{tikzpicture}

% \end{figure}
% \end{document}
    \caption{\textbf{Interest rate related model extensions.} We expand our re-implementation of DICE-2016-R with 1) a stochastic interest rate module with an optional stochastic abatement model, 2) the option to allow funding of abatement cost, and, 3) the option to apply non-linear discounting of damage cost. The latter can be used to penalize emission paths that generate cost overruns in terms of cost-per-GDP.}
    \label{fig:scheme_extensions}
\end{figure}

\bigskip

%%%%%%%%%%%%%%%%%%%%%%%%%%%%%%%%%%%%%%%%%%%%%%%%%%%%%%%%%%%%%%%%%%%%%%%%%%%%%%%%%%%%%%%%%%%%%%%%%%%
\clearpage
\section{Results}
\label{sec:Res}

\subsection{Intergenerational Inequality}

\begin{figure}[ht!]
    \centering
    \includegraphics[width=\textwidth]{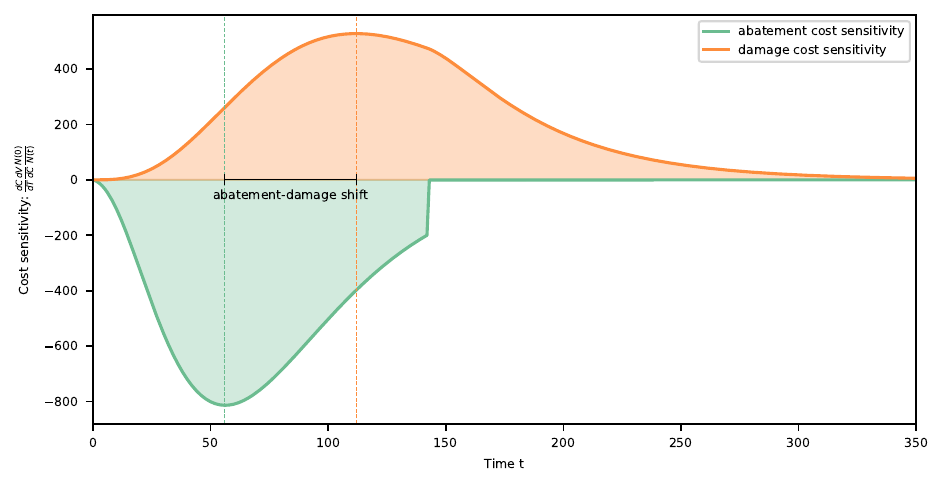}
    \caption[Temporal distribution of cost sensitivity]{
        The sensitivity $\frac{\mathrm{d}C_{\mathrm{A}}(t)}{\mathrm{d} T^{\mu=1}}$ and $\frac{\mathrm{d}C_{\mathrm{D}}(t)}{\mathrm{d}T^{\mu=1}}$ of abatement cost $C_{\mathrm{A}}$ and damage cost $C_{\mathrm{D}}$ on changes of the abatement policy $T^{\mu=1}$. The optimization is balancing the two areas under/above these curves - which is ultimately \emph{not} resulting in an intergenerational balancing of cost. The two sensitivities come with a significant time shift.
    }
    \label{fig:abatement_damage_sensitivity_weighted}
\end{figure}

The optimality condition of the IAM's emission path balances the sensitivity of the two cost components, abatement cost and damage cost, to policy changes (see~\cref{eq:cost_objective_mus:reduced_model}). We individually depict their temporal distribution: the weighted marginal change in abatement and damage cost with respect to a policy change in \cref{fig:abatement_damage_sensitivity_weighted} and find that an unconstrained IAM determines the optimality of the emission path by balancing the near-time abatement cost \emph{changes} with the future time damage cost \emph{changes}.
This may already constitute intergenerational inequality, as within that optimization the absolute burden of near-time abatement cost may be much lower than future-time damage cost.

\medskip

\begin{figure}[ht!]
    \centering
    \includegraphics[width=\textwidth]{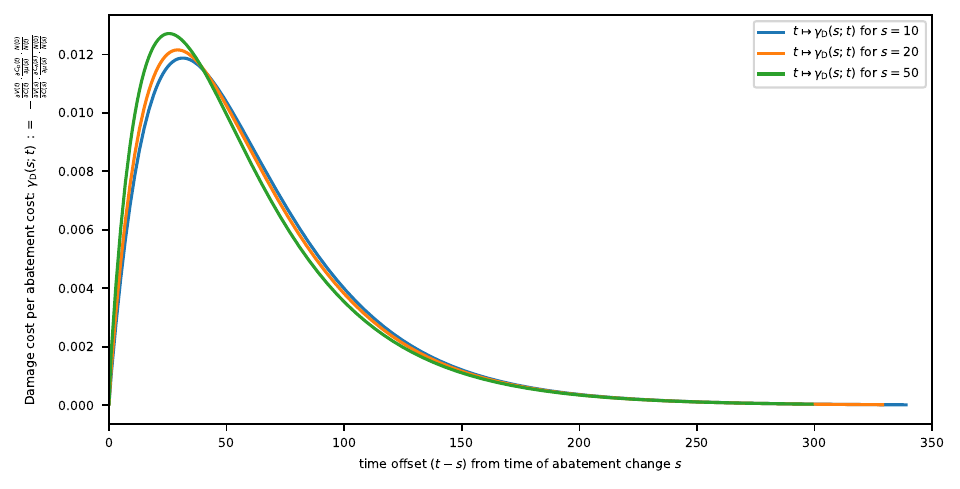}
    \caption{
        The density $t \mapsto \gamma_{\mathrm{D}}(s;t)$ for $s = 10, 20, 50$.
    }
    \label{fig:damage_cost_per_abatement_cost_denstiy}
\end{figure}

As changes in the abatement policy at time $s$ affect the abatement cost at time $s$ but damage cost at all future times $t \geq s$, we define $\gamma_{\mathrm{D}}(s;t)$ as \textit{the damage cost in $t$ per abatement cost in $s$ induced by a change in the abatement policy $\mu(s)$}, see~\eqref{eq:dice:desitySensiOfCostToAbetementValue}. We show that at the optimal emission path $\gamma_{\mathrm{D}}(s;t)$ is a probability distribution; a proof can be found in Supplement Section~\ref{sec:itergeneq:appendix:proof:costdensity}. This allows us to define the expected time-delay between an abatement and its effect on future damage cost, and we are able to determine it numerically.
\cref{fig:damage_cost_per_abatement_cost_denstiy} depicts the density $\gamma_{\mathrm{D}}(s)$.
We find that in the classical DICE 2016 model the expected time of damage cost per avoided abatement cost is approximately 70 years after the time of abatement. This observation agrees with \cref{fig:abatement_damage_sensitivity_weighted}, indicating that balancing marginal abatement and damage cost is a matter of intergenerational equity. We depict the temporal distribution of the total cost in \cref{fig:pathways_stochastic}, \cref{fig:pathways_panel_cost}.

\medskip

\subsection{Cost per GDP}

Since the optimal emission path balances marginal cost changes only but does not consider cost levels, we like to analyze the temporal distribution of the cost.
% Remark: The following paragraph is repeated in methods - may be shortened there.
To measure the burden borne by a generation, we consider the total cost per GDP. Putting nominal values relative to the GDP is a common measure, for example, to assess national debt \cite{daniel_pushing_2013}.
Here, it reflects that a wealthier or more productive generation can carry a greater burden, mimicking an effort-sharing scheme, and also considering the greater capacity of future generations under the expected baseline growth in DICE.

\subsection{Including stochastic interest rates into DICE}

Discounting plays a fundamental role in the linkage of present and future cost. We assess how uncertainty in discount rates alters the distribution of cost in the optimal emission paths.

We use a classical Hull-White model to model uncertainty in the rate of time preference $\rho$.

If interest rates are stochastic but the utility function remains deterministic,
adding stochastic interest rates does not change the interaction with the IAM, and hence the optimal emission path. However, this changes if one assumes that the abatement policy
is adapted to the (stochastic) changes in the interest rate level.

We allow for a stochastic abatement policy that adapts to the interest rate level, using the parametric model~\eqref{eq:stochastic_abatement_model_linear}. This reflects the possibility that the abatement policy can be adjusted to the interest rate scenarios.

With the stochastic abatement policy, we see an adverse effect: the expected abatement is smaller, the expected emissions are higher, and the expected total cost increases significantly for later years. Since by restricting a parameter of the stochastic abatement policy ($a_{1} = 0$ in \eqref{eq:stochastic_abatement_model_linear}) it agrees with the deterministic abatement model~\eqref{eq:abatementModel:determ:oneparam}, it is apparent that the effect is created solely by allowing the abatement speed to depend on the interest rate level (i.e., $a_{1}$ being a free parameter in \eqref{eq:stochastic_abatement_model_linear}).

\begin{figure}[ht!]
    \centering
    \includegraphics[width=\textwidth]{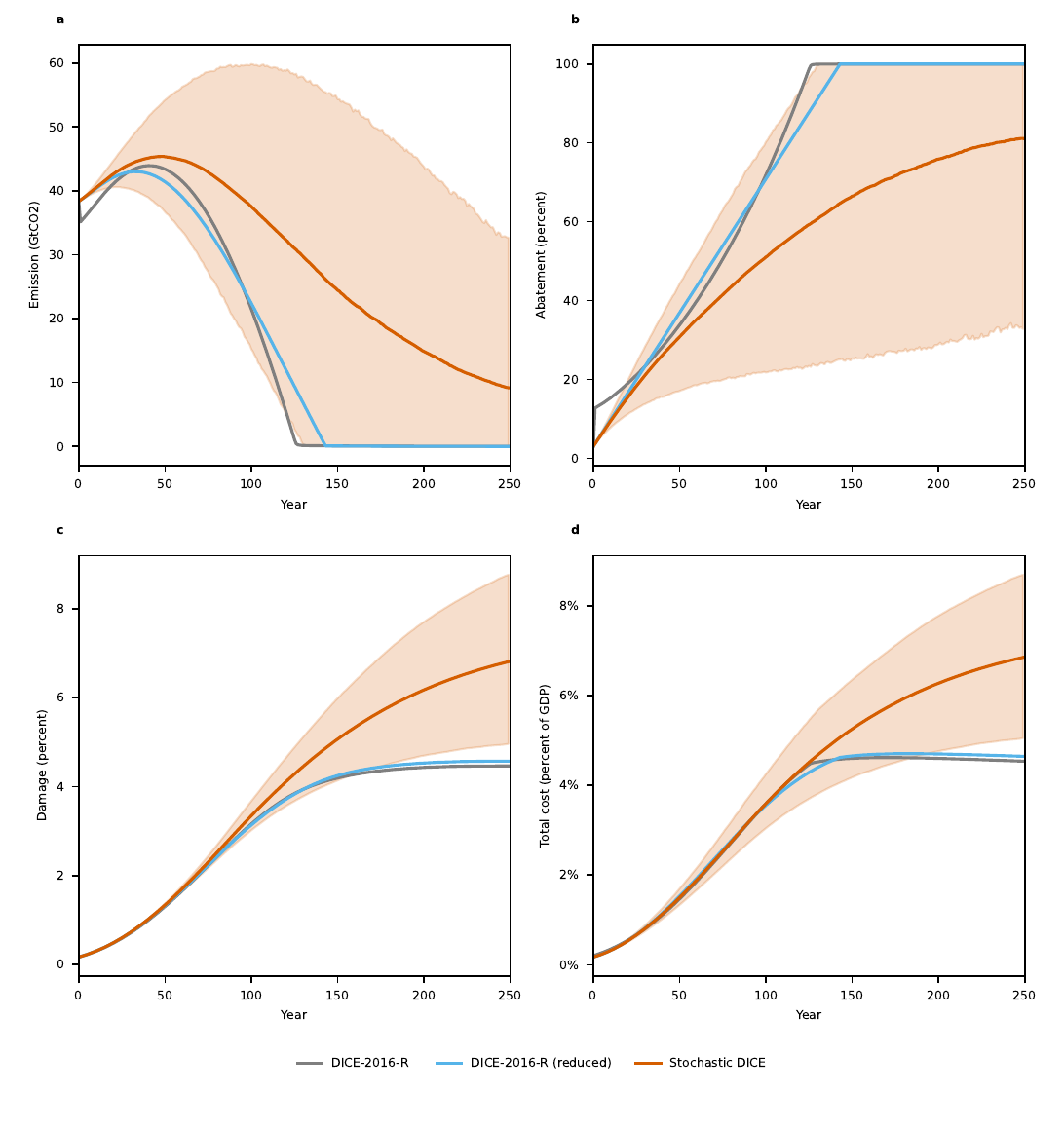}
    \caption{
        Classical DICE (grey), Classical DICE with reduced Abatement Model \eqref{eq:abatementModel:determ:oneparam} (blue) and Stochastic DICE Model (red). Restricting the abatement policy to the 1-parameter model \eqref{eq:abatementModel:determ:oneparam} has a comparably small effect on the emission pathways. Allowing adaptation of the abatement policy to stochastic interest rates increases intergenerational inequality \emph{and} exhibits a significant risk of even higher intergenerational inequality
        (\textbf{a}: emissions, \textbf{b}: abatement, \textbf{c}: damages, and \textbf{d}: total GDP relative cost). Shaded area shows the 10th to 90th~percentiles.
        The deterministic model (blue) uses the certainty‑equivalent discount rate $\bar{r}$ from \eqref{eq:dice:stochastic_rates_in_deterministic_model}; therefore, differences in the red line arise solely from interest‑rate–driven policy adaptation ($a_{1} \neq 0$) and damage nonlinearity, not from a change in average discounting.
    }
    \label{fig:pathways_stochastic}
\end{figure}

\cref{fig:pathways_stochastic} depicts the results of stochastic interest rates, both in expectation and in a percentile measure (VaR), on emissions, optimal abatement, damage, and total cost-per-GDP.

This shows that under the stochastic interest rate and stochastic abatement model, the optimal policy will increase intergenerational inequality and, in addition, exhibit the risk of further increases in intergenerational inequality (as given by the percentiles). In other words, the DICE model exhibits convexity with respect to the interest rates.

This effect is qualitatively independent of the interest rate level. The strength of the effect increases with an increase in interest rate volatility, \cref{fig:pathways_stochastic_variation}.
For the limit case of vanishing interest rate volatility $\sigma = 0$, we recover the classical deterministic DICE model.

\subsection{Improving Intergenerational Equity}

We investigate the effect of the two model extensions ``funding of abatement cost'' (Section~\ref{sec:diceExtension:fundingOfAbatement}) and ``non-linear discounting of damage cost'' (Section~\ref{sec:nonlinear_discounting}) on the distribution of cost in the deterministic and stochastic model, i.e., how they improve cost-related intergenerational equity.

The parameter choices are exemplary; other experiments with similar results can be reproduced in the published source code.

\subsubsection{Funding of Abatement}

We allow funding of abatement costs to reduce the time lag between abatement cost and damage cost. As we account for interest for the funding, this does not alter the effect of discounting but brings the two components to similar economical regions.

Shifting cost via a loan to the future may be considered adverse to intergenerational equity; however, our numerical analysis indicates that it is beneficial as it incentivizes higher abatement.

In a numerical experiment, with a funding period of 60 years with equal-value annual repayment (including interest), funding of abatement reduces the cost in the very far future without substantial increase in the near future (\cref{fig:pathways_panel_cost}, left). An even more notable contribution is that it significantly reduces the cost risk if discount rates are stochastic (\cref{fig:pathways_panel_cost}, right).

However, with respect to intergenerational equity, improvements are somewhat unbalanced as a funding of abatement still leaves a high burden for near future generations.

\subsubsection{Funding Risks of Damage}

We add a non-linear discounting of damages. This model extension allows the discount factor to depend on the magnitude of the cash flow, mimicking that funding of larger cash flows requires a larger premium to compensate for default risk or other frictions, a \textit{default compensator}. As such, larger damages get a larger weight; consequently, the abatement policy optimization will prefer emissions paths that avoid very large damages, which is beneficial with respect to intergenerational equity.

\begin{figure}[hbtp]
    \centering
    \includegraphics[width=\textwidth]{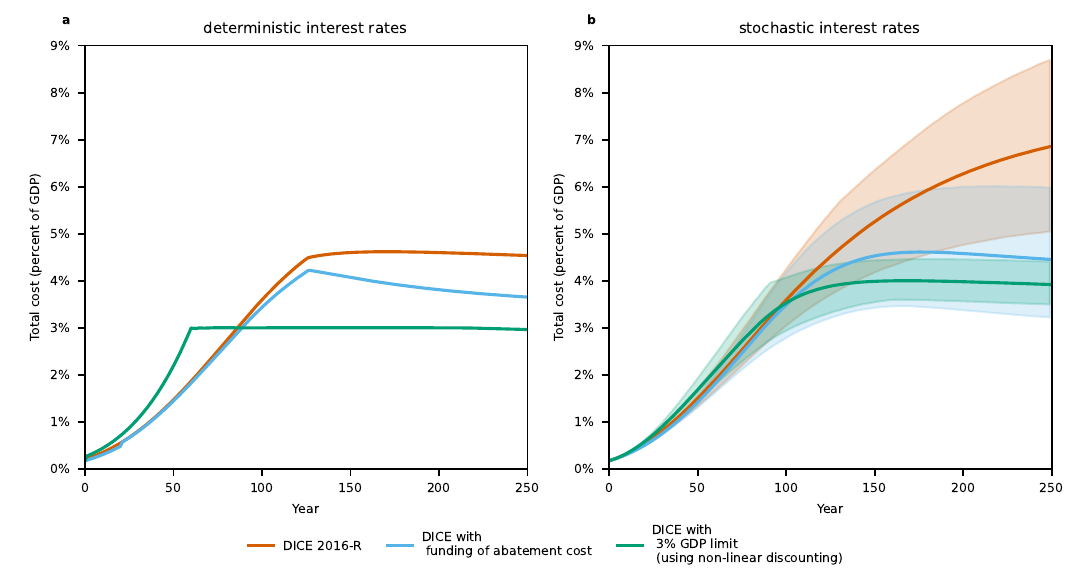}
    \caption[The cost-per-GDP with and without stochastic interest rates and different model extensions]{The cost-per-GDP in \textbf{a} the model with deterministic interest rates and \textbf{b} the model with stochastic interest rates, for the otherwise unmodified model (red) and the model with the extension ``funding of abatement cost'' (blue) and ``GDP limited non-linear discounting'' (green). Shaded area shows the 10th to 90th percentiles.
	}
    \label{fig:pathways_panel_cost}
\end{figure}

In our numerical experiment, we consider a significant default compensator for damage cost over 3\% of the GDP, effectively limiting cost of optimal emission paths to $3 \%$. We then find strong improvements to intergenerational equity in the deterministic DICE model and the DICE model with a stochastic interest rate, as shown in~\cref{fig:pathways_panel_cost}. Intergenerational equity is improved with respect to the expectation and the uncertainty in the respective optimal emission pathway, represented by the 10th and 90th~percentiles. The associated model quantities (emission, abatement and damage) are depicted in \Cref{fig:pathways_deterministic_with_extension,fig:pathways_stochastic_with_extension}.

Due to the limits of scaling up abatement immediately and the inherent preference for current generations in the DICE model, the present generations still face the lowest costs.

\medskip

\begin{figure}[ht!]
    \centering
    \includegraphics[width=\textwidth]{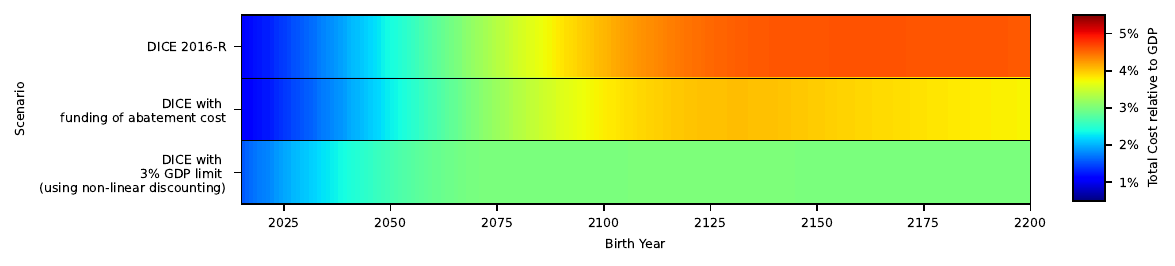}
    \caption[Illustration of generational inequality, deterministic model]{\textbf{Illustration of generational inequality, deterministic model:} Lifetime-average cost relative to GDP by year of birth in the different model specifications up to 2200. Normalized by annual GDP and scaled such that blue means lower, red higher average total cost values than 3\% of GDP for this generation.}
    \label{fig:generational_burden_deterministic}
\end{figure}

\begin{figure}[ht!]
    \centering
    \includegraphics[width=\textwidth]{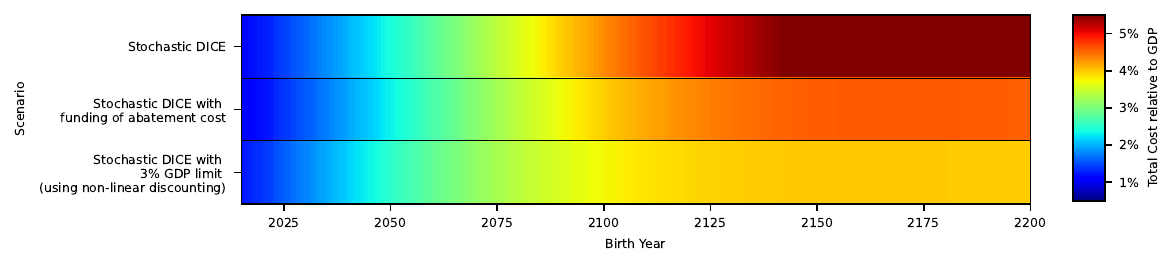}
    \caption[Illustration of generational inequality, stochastic model]{\textbf{Illustration of generational inequality, stochastic model:} Lifetime-average cost relative to GDP by year of birth in the different model specifications up to 2200. Normalized by annual GDP and scaled such that blue means lower, red higher average total cost values than 3\% of GDP for this generation.}
    \label{fig:generational_burden_stochastic}
\end{figure}

%\cref{fig:pathways_panel_cost} depicts the instantaneous cost-per-GDP ($C(t)/GDP(t)$).
For an assessment of the generational burdens we consider the lifetime-average value using population projections from 2015 to 2100 \cite{noauthor_world_nodate} to calculate burdens experienced during each birth-year cohorts projected lifetime.
We measure the burden in percentage-of-GDP (\cref{sec:costPerGDP}), where we choose 3\% as the neutral level (green), \cite{Fries2024FairShareOfGDP}. While differences are small for the time covered by population projections (\cref{fig:generational_burden_deterministic} and \cref{fig:generational_burden_stochastic}), extrapolation using the fixed 2100 life expectancy shows a clear divergence at the end of the century. Notably, the stochastic variant does not stay exactly below 3\% of GDP on average since high-risk scenarios are not fully offset. The deterministic model keeps the 3\%-limit and illustrates the equal burden between generations enabled by imposing GDP relative cost limits.

%%%%%%%%%%%%%%%%%%%%%%%%%%%%%%%%%%%%%%%%%%%%%%%%%%%%%%%%%%%%%%%%%%%%%%%%%%%%%%%%%%%%%%%%%%%%%%%%%%%
\clearpage
\section{Discussion} % target 500 words 
\label{sec:dice_extentions:discussion}

We analyzed the DICE integrated assessment model and derived metrics to assess the temporal cost structure and thus intergenerational equity of the optimal emission pathway. We extend the DICE integrated assessment model and compare several extensions for their effects on intergenerational equity. Our methods and extensions apply to integrated assessment models in general and can also be extended to the benefit-cost analysis of climate mitigation.

\medskip

%results
Our first observation is that the DICE model optimization results in an unequal temporal distribution of policy costs measured in terms of cost-per-GDP. This is because the optimization does not consider any (cost-related) intergenerational equity but balances the marginal aggregated cost changes induced by a policy change, while absolute cost and its distribution over time are irrelevant to the objective function.

Considering uncertainty in the rate of time-preference and thus overall interest rates, this effect is amplified when the stochastic adaptation of the abatement strategy generates a significant risk of increased cost. This indicates that repeated re-calculation of the DICE model to current interest rate levels amplifies intergenerational inequality.

We consider natural model extensions by allowing the funding of abatement cost and non-linear discounting of damage cost.
Funding of abatement cost may be considered as modeling the backing of abatement costs by government loans. We find that this may have a positive effect on intergenerational equity in terms of the cost distribution as it improves the alignment of abatement and damage cost.
The non-linear discounting of damage cost may be considered as modeling financing risk of damage costs, where very large cost have higher financing cost. The non-linear damage cost can be used as a constraint that allows to effectively limit the generational cost to a specific percentage of the GDP. 

%caveats
Using a simple IAM includes several caveats. First, the representation of damages could be improved \cite{Burke2015,Glanemann2020,Hansel2020}. Second, the lack of regional resolution compared to, e.g., econometric damage estimates \cite{kotz_day--day_2021,kotz_effect_2022} ignores regional inequalities in interest rate dynamics, exposure, and potential distributional effects of projected costs \cite{andreoni_inequality_2023}. Since our study aims to introduce methods such as limiting total costs of climate change by a proportion of GDP to the debate on mitigation pathways, these limitations could be addressed by including the presented concepts in more complex assessments of mitigation pathways \cite{rennert_comprehensive_2022,van_der_wijst_new_2023}.

%summary
Modifying the DICE model in a modular way, we address the issue that the classical model does not consider uncertainty in interest rates and its effects on equity between generations. The extensions presented accounting for the funding cost of abatement and increased financing cost of large damages reduce intergenerational inequality. This is amplified when considering stochastic interest rate risk due to convexity induced by the DICE objective function. Constraining the total costs of mitigation and damages relative to GDP could be used as a simple objective to ensure intergenerational equity of climate mitigation.
Our results can also be interpreted as supporting the design of financing mechanisms such as a climate transformation fund that channels dedicated, long-term resources into mitigation and adaptation, see also~\cite{Fries2023CO2}, which builds on the modeling framework developed in the present work and verifies that a climate transformation fund can be funded with less than 3 \% of the GDP. Implementing these mechanisms in practice would require collaboration between policymakers, financial institutions, and affected communities to ensure that they are economically sustainable, politically viable, and perceived as fair between generations.

%%%%%%%%%%%%%%%%%%%%%%%%%%%%%%%%%%%%%%%%%%%%%%%%%%%%%%%%%%%%%%%%%%%%%%%%%%%%%%%%%%%%%%%%%%%%%%%%%%%
\clearpage
\section{Methods}

The starting point of our modeling exercise is DICE-2016 \cite{Nordhaus2013, Nordhaus2017}.
\begin{comment}
The DICE-2023 description \cite{barrage_policies_2024} was only published parallel to this research and differences are of minor relevance to our model extensions. 
\end{comment}

Except where otherwise stated, we use the same model components as in DICE-2016. A detailed description of our implementation of the DICE model and its extensions can be found in~\cite{fries_intergenerational_2023}. We use the same notation, except that we denote the cost functions by $C$.\footnote{In \cite{Nordhaus2017} the letter $C$ was used for consumption, which we do not refer to here.}

We redefine the objective function to be suitable for stochastic model extensions; see Section~\ref{sec:model_calibration}.
% LINK to Appendix?
% Discuss other work?
%
We then analyze the temporal distribution of cost and the cost sensitivity to policy changes in the classical DICE model in Section~\ref{sec:introMetrics} as an indicator of intergenerational equity.
Section~\ref{sec:diceExtension:deferredfinancing} introduces model extensions, which allow to alter the distribution of cost towards intergenerational equitable climate mitigation. Most importantly, we allow for a stochastic time preference $\rho$ modeled by a classical interest rate model, as described in \cref{sec:introducingStochInterestRates}. 

\subsection{Model Objective}
\label{sec:model_calibration}

The objective of the DICE model is to determine the abatement rate $t \mapsto \mu(t)$ and the savings rate $t \mapsto s(t)$ that maximize a given objective function.

The objective function of the classical DICE model is the integrated discounted welfare
\begin{equation}
	\label{eq:integrated_discounted_welfare}
	\int_{0}^{T} V(t) \cdot \frac{N(0)}{N(t)} \mathrm{d}t \text{.}
\end{equation}
Here, $V$ denotes the utility and $N$ denotes the numéraire. DICE-2016 is recovered for $N(t) = \exp(\rho \cdot t)$. We use the more general notion of a numéraire, as we will consider a more general model later. For the stochastic model, the objective function will be an expectation of \eqref{eq:integrated_discounted_welfare} and a quantile for uncertainty analysis.

\subsubsection{Abatement Policy - Classical DICE Model and Reduced Model}

In the time-discrete classical DICE model, optimization determines the optimal parameters $i \mapsto \left(\mu(t_{i}), s(t_{i})\right)$.

% \subsubsection{Parametric Abatement Policy - Reduced Model}

For comparison with a parametric \emph{stochastic} abatement model, which we introduce later as part of our model extensions, we define a parametric abatement model
\begin{equation}
    \label{eq:abatementModel:determ:oneparam}
    \mu(t) \ = \ \min \left( \mu(0) + \frac{1 - \mu(0)}{T^{\mu=1}} \ t , 1.0 \right) \text{,}
\end{equation}
where the parameter $T^{\mu=1}$ represents the time required to reach 100\% abatement. This model has only a single parameter, however it turns out that the optimal abatement policy in the full parameter model is comparably close to a functional form of the type \eqref{eq:abatementModel:determ:oneparam}.

\subsection{Metrics Describing Intergenerational Burden}
\label{sec:introMetrics}

To characterize the resulting optimal mitigation paths with respect to Intergenerational equity, we consider the total cost.
% XXX TODO: Add reference to SCC in SI

\subsubsection{Cost}

We analyze the temporal distribution of cost $t \mapsto C(t)$, which is the sum of the abatement cost $C_{\mathrm{A}}$ and the damage cost $C_{\mathrm{D}}$
\begin{equation*}
	C(t) \ := \ C_{\mathrm{A}}(t) + C_{\mathrm{D}}(t) \text{,}
\end{equation*}
and investigate the interplay of abatement and damage cost.

\subsubsection{Cost per GDP}
\label{sec:costPerGDP}

% XXX TODO Paragraph slightly repeated in results

We measure the burden borne by a generation by considering the total cost per GDP, as it is common in other contexts, for example, to assess national debt \cite{daniel_pushing_2013}.
Here, it can be considered as mimicking an effort-sharing scheme, that a more wealthy or productive generation can carry a greater burden,

To allow for a netting within a generation, one may consider the (discounted) average value over a generation lifetime. For a given time $s$ let $T^{\mathrm{L}}(s)$ denote the expected lifetime of a generation. We then consider the lifetime-average cost and GDP,
\begin{equation*}
	\overline{C}(s) \ := \ \int_{s}^{s+T^{\mathrm{L}}(s)} C(t) \frac{N(s)}{N(t)} \mathrm{d}t \text{,}
	\qquad
	\overline{GDP}(s) \ := \ \int_{s}^{s+T^{\mathrm{L}}(s)} GDP(t) \frac{N(s)}{N(t)} \mathrm{d}t \text{.}
\end{equation*}
In our numerical experiments, we measure the instantaneous cost-per-GDP $\frac{C(s)}{GDP(s)}$ as well as the lifetime-average cost-per-GDP $\frac{\overline{C}(s)}{\overline{GDP}(s)}$.

\subsubsection{Sensitivity of Damage and Abatement Cost to Policy Changes}

The following shows why the unconstrained optimal emission path may result in an unequal distribution of the total cost and the total cost per GDP.

The optimal abatement policy $\mu$, representing the equilibrium state of the model, fulfills (differentiating the objective function \eqref{eq:integrated_discounted_welfare} with respect to the abatement policy $\mu$)
\begin{align}
    \frac{\mathrm{d}}{\mathrm{d} \mu} \ \int_0^{T} V(t) \frac{N(0)}{N(t)} \mathrm{d}t
    & \ = \ \int_{0}^{T} \frac{\mathrm{d}V(t)}{\mathrm{d}C(t)} \left( \frac{\mathrm{d}C_{\mathrm{D}}(t)}{\mathrm{d} \mu} + \frac{\mathrm{d}C_{\mathrm{A}}(t)}{\mathrm{d} \mu} \right) \ \frac{N(0)}{N(t)} \mathrm{d}t \ \stackrel{!}{=} \ 0
    \label{eq:cost_objective}
\end{align}
As the objective function depends on the cost, two transformations take place. First, the value $V$ is defined as utility. Because the utility function is concave, the weight $\frac{\partial V(t)}{\partial C(t)}$ decays. Second, the values are discounted. With a constant positive time-preference rate, the discount factor constitutes an exponentially decaying weight $\frac{N(0)}{N(t)}$ applied to the cost. The product of both weights represents the \textit{cost-to-value-weight}.

From \cref{eq:cost_objective}, we see that the model optimization may result in some intergenerational inequality: Let $s$ be a fixed time at which a change in the abatement policy $\mu(s)$ is considered. Because $\mathrm{d}C_{\mathrm{D}} / \mathrm{d} \mu(s) < 0$ and $\mathrm{d}C_{\mathrm{A}} / \mathrm{d} \mu(s) > 0$ we see from \cref{eq:cost_objective} that the model equates the \emph{decrement} in the cost of damage to the \emph{increment} in the cost of abatement, both weighted by the value-to-cost-sensitivity $\mathrm{d}V(t) / \mathrm{d}C(t) \cdot N(0)/N(t)$. We have
\begin{equation}
    \label{eq:cost_objective_mus}
    \int_{0}^{T} \frac{\mathrm{d}V(t)}{\mathrm{d}C(t)} \frac{\mathrm{d}C_{\mathrm{D}}(t)}{\mathrm{d} \mu(s)} \ \frac{N(0)}{N(t)} \mathrm{d}t
    \ \stackrel{!}{=} \
	- \int_{0}^{T} \frac{\mathrm{d}V(t)}{\mathrm{d}C(t)} \frac{\mathrm{d}C_{\mathrm{A}}(t)}{\mathrm{d} \mu(s)} \ \frac{N(0)}{N(t)} \mathrm{d}t \text{,}
\end{equation}
where $t \mapsto \frac{\mathrm{d}C_{\mathrm{A}}(t)}{\mathrm{d} \mu(s)}$ and $t \mapsto \frac{\mathrm{d}C_{\mathrm{D}}(t)}{\mathrm{d} \mu(s)}$ exhibit different temporal distributions.

The sensitivity $\frac{\mathrm{d}}{\mathrm{d} \mu}$ is a functional derivative or gradient. Given the one-parameter abatement model \eqref{eq:abatementModel:determ:oneparam} the equilibrium state is given by
\begin{equation}
    \label{eq:cost_objective_mus:reduced_model}
    \int_{0}^{T} \frac{\mathrm{d}V(t)}{\mathrm{d}C(t)} \frac{\mathrm{d}C_{\mathrm{D}}(t)}{\mathrm{d} T^{\mu=1}} \ \frac{N(0)}{N(t)} \mathrm{d}t
    \ \stackrel{!}{=} \
	- \int_{0}^{T} \frac{\mathrm{d}V(t)}{\mathrm{d}C(t)} \frac{\mathrm{d}C_{\mathrm{A}}(t)}{\mathrm{d} T^{\mu=1}} \ \frac{N(0)}{N(t)} \mathrm{d}t \text{,}
\end{equation}
and we can plot the (weighted) damage cost sensitivity $\frac{\mathrm{d}C_{\mathrm{D}}(t)}{\mathrm{d} T^{\mu=1}}$ and the (weighted) abatement cost sensitivity $\frac{\mathrm{d}C_{\mathrm{A}}(t)}{\mathrm{d} T^{\mu=1}}$, see \cref{fig:abatement_damage_sensitivity_weighted}.

\subsubsection{Damage Cost per Abatement Cost upon Abatement Policy Change}

% XXX TODO this repeated in results
A change in the abatement policy at time $s$ affects the abatement cost $C_{\mathrm{A}}(s)$ only at time $s$, whereas it affects damage cost $C_{\mathrm{D}}(t)$ for all $t \geq s$.
This motivates to consider the damage cost per abatement cost induced by a change in the abatement policy $\mu(s)$, i.e.,
\begin{equation*}
    \frac{\partial C_{\mathrm{D}}(t)}{\partial \mu(s)} \big/ \frac{\partial C_{\mathrm{A}}(s)}{\partial \mu(s)} \text{.}
\end{equation*}
As the model optimization weights cost by the cost-to-value weight, we consider the corresponding weighted damage cost per abatement cost sensitivity.

\paragraph{Definition (Damage Cost per Abatement Cost upon Abatement Policy Change)}

We define\footnote{The minus in \eqref{eq:dice:desitySensiOfCostToAbetementValue} accounts for the fact that abatement cost sensitivity and damage cost sensitivity have opposite signs.}
\begin{equation}
    \label{eq:dice:desitySensiOfCostToAbetementValue}
    \gamma_{\mathrm{D}}(s;t) \ := \ - \frac{\frac{\partial V(t)}{\partial C(t)} \cdot \frac{\partial C_{\mathrm{D}}(t)}{\partial \mu(s)} \cdot \frac{N(0)}{N(t)}}{\frac{\partial V(s)}{\partial C(s)} \cdot \frac{\partial C_{\mathrm{A}}(s)}{\partial \mu(s)} \cdot \frac{N(0)}{N(s)}} \text{.}
\end{equation}

\paragraph{Lemma}

At the optimal abatement policy $\mu$, for fixed $s$ the integral over $t \mapsto \gamma_{\mathrm{D}}(s;t)$ equals $1$, i.e., $\gamma_{\mathrm{D}}(s)$ can be interpreted as a density, and we may interpret
\begin{equation*}
    t_{D}(s) \ := \ \int_{0}^{\infty} t \cdot \gamma_{\mathrm{D}}(s;t) \ \mathrm{d}t \text{.} 
\end{equation*}
as the expected time of damage cost per avoided abatement cost at time $s$.

\bigskip

The proof is provided in the Appendix~\ref{sec:itergeneq:appendix:proof:costdensity}.

\bigskip

Our numerical analysis will show a significant (expected) time lag for $t_{D}(s)$ (roughly 70 years in a DICE-2016). This means that \textit{marginal} gains of one generation are equated to \textit{marginal} losses of another generation. The optimization does not equalize the burden but rather the weighted marginal burden. In this sense, the optimization is indifferent to the temporal distribution of the absolute burden.

\begin{comment}
%removed
The conditions \eqref{eq:cost_objective_mus} and \eqref{eq:cost_objective_mus:reduced_model} are plausible if a single agent optimizes its cost but may be problematic if the costs are distributed between different generations. Note that the two individual contributions in~\cref{eq:cost_objective_mus:reduced_model} occur at different times. %is this a bit convoulted? maybe different times is enough?    
\end{comment}

% TODO Reference section with figure fig:abatement_damage_sensitivity_weighted below

% TODO XXX Just have convexity as a small remark above
\begin{comment}
    
\subsection{Convexity}
\label{sec:convexity}
Our numerical analysis shows that a continuous adaptation of the abatement strategy may increase the risk and expectation of intergenerational inequality measures as cost-per-GDP. The expected emissions/damage/cost are higher than the emissions/damage/cost calculated for the expected interest rate level.

Such an effect may come as a surprise. However, it is well known from mathematical finance, known under the name \emph{convexity}. The effect corresponds to Jensen's inequality, that the expectation of a convex function of a random variable is higher than the convex function applies to the expectation of the random variable.

Here, the random variable is the exogenous given discount rate, or stochastic process of the interest rates, and the integrated assessment model acts as a convex function. In other words: DICE has convexity in the interest rates once the abatement strategy depends on the interest rate level.

\end{comment}

\subsection{Model Extensions}
\label{sec:diceExtension:deferredfinancing}

We present modifications to integrated assessment models to investigate the effects of stochastic interest rates, funding abatement costs, and non-linear financing costs \cite{fries_non-linear_2021}. These may also be incorporated into more complex IAMs. We use the DICE model to illustrate potential effects.

\subsubsection{Stochastic Interest Rates}
\label{sec:introducingStochInterestRates}

In classical IAMs, the time preference rate $\rho$ is typically a time-dependent deterministic function, which is often a constant.

We model a stochastic interest rate (time-preference rate) $\rho$. The model provides the stochastic numéraire $N$ (a bank account accruing continuously at rate $\rho$) as well as the stochastic forward rate $FR$. Setting the volatility of the interest rate to zero recovers the deterministic model.

Our implementation allows for the use of a general discrete forward rate model (LIBOR Market Model). Our results were obtained using the classical Hull-White model \cite{hull1990pricing, FriesLectureNotes2007}, that is
\begin{equation}
	\label{eq:dice:hull_white_model}
	\begin{split}
		\mathrm{d} \rho(t) & \ = \ ( \theta(t) - a(t) \rho(t) ) \ \mathrm{d} t \ + \ \sigma(t) \ \mathrm{d} W(t) \text{,} \quad \rho(t_{0}) \ := \ r_{0} \text{,} \\
		\mathrm{d} N(t) & \ = \ \rho(t) \ N(t)  \ \mathrm{d} t \text{,}
	\end{split}
\end{equation}
where an exact time-discretization scheme \cite{fries_hull_white2016} was used. Our stochastic interest rate model is parameterized such that the optimal \emph{deterministic} abatement strategy would result in the same optimal emission pathway as the model with deterministic interest rates.

We set $r_{0}$, $\theta(t)$ so that the certainty‑equivalent rate $\bar{\rho}(t)$ equals the deterministic time preference rate $\rho$, such that outcomes are exactly identical to the deterministic model \eqref{eq:dice:stochastic_rates_in_deterministic_model}. Mean reversion and volatility were chosen to reflect plausible long‑term interest‑rate dynamics. The chosen mean‑reversion speed of $a(t) = 0.02$ implies a half‑life of $\ln(2)/0.02 \approx 35$ years, while the chosen volatility of $\sigma(t) = 0.3 \%$ produces a dispersion of future rates that remains within a realistic range—roughly $\pm 3$ percentage points around the long‑run mean over a century.

While our result should be interpreted as illustrative rather than predictive, showing that stochastic interest rates generate adverse effects, the interest rate scenarios generated by our parameter choice are realistic, rather conservative: the 1st and 99th~percentile of the generated interest rates at t = 100 years are -1~\% and 5~\%, respectively. For a corresponding visualization, see~\cite{fries_intergenerational_2023}.

\bigskip
%\subsubsection{Objective Function for the Stochastic Model}

As we consider a model extension where $N$ and $V$ are stochastic processes, we redefine the objective function \eqref{eq:integrated_discounted_welfare} as a random variable. For the optimization we consider the expected welfare
\begin{equation}
	\label{eq:integrated_discounted_welfare:expectation}
	\mathrm{E} \left( \int_{0}^{T} V(t) \cdot \frac{N(0)}{N(t)} \mathrm{d}t \right) \text{.}
\end{equation}

\begin{comment}    
For uncertainty analysis, we plot a risk measure, here the value-at-risk risk, that is, a percentile of the stochastic welfare,
\begin{equation}
	\label{eq:integrated_discounted_welfare:risk}
	\mathrm{VaR}_{\alpha} \left( \int_{0}^{T} V(t) \cdot \frac{N(0)}{N(t)} \mathrm{d}t \right) \text{.}
\end{equation}
\end{comment}

%%%%%%%%%%%%%%%%%%%%%%%%%%%%%%%%%%%%%%%%%%%%%%%%%%%%%%%%%%%%%%%%%%%%%%%%%%%%%%%%%%%%%%%%%%%%%%%%%%%

%\subsubsection{Interaction of Stochastic Discount Rates with the IAM}

If interest rates are stochastic but the utility function $V(t)$ remains deterministic, adding stochastic interest rates does not change the interaction with the IAM because
\begin{equation}
	\label{eq:dice:stochastic_rates_in_deterministic_model}
	\begin{split}
		\mathrm{E} \left( \int_{0}^{T} V(t) \frac{N(0)}{N(t)} \mathrm{d}t \right) & \ = \ \int_{0}^{T} V(t) \ \mathrm{E} \left( \frac{N(0)}{N(t)} \right) \mathrm{d}t \\
		& \ = \ \int_{0}^{T} V(t) \exp\left( - \int_{0}^{t} \bar{r}(s) \mathrm{d}s \right) \mathrm{d}t \text{,}
\end{split}
\end{equation}
where $\bar{r}(t) \ = \ - \frac{\partial}{\partial t} \log\left( \mathrm{E} \left( \exp\left( - \int_{0}^{t} \rho(s) \mathrm{d}s \right) \right) \right)$.

Owing to the lack of feedback that allows adjustment of the abatement policy and resulting damages, adding stochastic interest rates does not yet introduce changes in the model dynamics. The model produces the same emission pathway as for the deterministic model.
However, this changes if one assumes that the abatement policy is adapted to the (stochastic) changes in the interest rate level.

\begin{comment}
\footnote{Our choice to make interest rates stochastic through uncertain time preference rates is only exemplary. The interest rate is an important factor in linking present abatement cost to the avoided future damage cost. While one may introduce stochasticity in many other state variables, the use of an abatement policy adapted to the economic factors (interest rates) will already lead to all other state variables becoming stochastic.}    
\end{comment}

\subsubsection{Stochastic Abatement Policy}

Assuming that a change in the interest rates triggers a re-calculation of the model optimal abatement policy, a stochastic interest rate implies a stochastic abatement policy. The abatement policy $t \mapsto \mu(t)$ becomes a stochastic process that adapts to changes in interest rates. This reflects the possibility that the abatement policy can be adjusted to the interest rate scenarios.

Determining an optimal stochastic abatement policy is a classical optimal exercise problem, as in the valuation of financial products with early exercise options in mathematical finance. Here, $\mu$ takes the role of the exercise strategy. As the time-$t$ optimal exercise is required to be a $\mathcal{F}_{t}$-measurable, it may be represented as a functional form of time-$t$ measurable random variables \cite{FriesLectureNotes2007}.

An approach to determine the optimal exercise strategy is to parameterize $\mu(t)$ as a functional form of $\mathcal{F}_{t}$-measurable random variables and apply a global optimization. As we utilize a one-factor Markovian model for $\rho$, we consider a functional form where $\mu$ depends on $t$ and the current interest rate level $\rho(t)$.

To obtain a robust optimization, we propose the parametric stochastic abatement model
\begin{equation}
	\label{eq:stochastic_abatement_model_linear}
    \mu(t,\omega) = \min\left( \mu_{0} + \left( a_{0} + a_{1} \cdot \rho(t,\omega) \right) \cdot t , 1.0 \right) \text{,}
\end{equation}
where the abatement speed is a (linear) function of the interest rate level $\rho$.

The model can be seen as a first-order approximation of general policies. Its simplicity allows the interpretation of the parameters: $a_{0}$ is the abatement speed at the zero interest rate level and $a_{1}$ describes the change in the abatement speed per interest rate change,
\begin{equation*}
    a_{0} \ = \ \frac{\partial \mu(t)}{\partial t}\big\vert_{r = 0} \text{,} \qquad a_{1} \ = \ \frac{\partial}{\partial r} \ \frac{\partial \mu(t)}{\partial t} \text{.}
\end{equation*}
\begin{comment}    
Obviously, the stochastic abatement models may achieve a larger-or-equal welfare or lower-or-equal welfare-risk than the corresponding deterministic model \eqref{eq:abatementModel:determ:oneparam} simply due to the additional degree of freedom $a_{1}$ in the abatement optimization.
\end{comment}
The optimization of the model \eqref{eq:stochastic_abatement_model_linear} results in a negative value for $a_{1}$, reflecting that a higher interest rate corresponds to a slower abatement speed.

%tThe functional form \eqref{eq:stochastic_abatement_model_linear} is already rich enough to shed light on some stylized effects, e.g., exhibiting a convexity of the model that suggests that successive re-calculation of the deterministic model worsens Intergenerational equity. In the following, we consider the simple linear functional form \eqref{eq:stochastic_abatement_model_linear}. In \cite{fries_intergenerational_2023}, we also investigated a higher-order functional form that does not give qualitatively different results.

\subsubsection{Funding of Abatement Costs}
\label{sec:diceExtension:fundingOfAbatement}

The simple damage function of the DICE model provides damages as a reduction of global GDP. Damage may be reduced by abatement of emissions. Both abatement and damage are associated with costs. The time-$t_{i}$ costs are directly deduced from the time-$t_{i}$ GDP. The remainder is then available for consumption or investment. Investment adds to the capital, which determines the GDP of the next time $t_{i+1}$.
However, abatement cost and damage cost resulting from the same abatement policy decision occur at significantly different times, as our analysis shows.

To resolve the time-lag between abatement and damage costs, we introduce the option of funding abatement cost. As the abatement of emissions is a planned process of societal relevance, it is reasonable to assume that abatement costs are covered by a loan for which interest corresponds to the current discount rate. Our model allows the application of a funding spread, \ie higher interest rates for loans than for discounting, but this is not considered in this general introduction of the phenomenon. In future work, the concept of the climate beta could be used to introduce an interest rate spread for abatement investments comparable to the modifications in DICE-2023 \cite{barrage_policies_2024}.

We define $C_{\mu}(t)$ as the instantaneous abatement costs in time $t$, i.e., the quantity that was formerly denoted by $C_{\mathrm{A}}(t)$. We allow that $C_{\mu}(t)$ is funded for a period $\Delta T_{\mathrm{A}}$. Thus, the abatement costs of $C_{\mu}(t)$ are accrued with the forward rate $FR(t,t+\Delta T_{\mathrm{A}};t)$ observed in $t$, such that the realized abatement cost $C_{\mathrm{A}}$ at time $t+ \Delta T_{\mathrm{A}}$ is
\begin{equation}
    \label{eq:dice:fundingOfAbatementCost}
    C_{\mathrm{A}}(t + \Delta T_{A}) \ := \ C_{\mu}(t) \ \left( 1 + FR(t,t+\Delta T_{A};t) \Delta T_{\mathrm{A}} \right) \text{.}
\end{equation}
For the case of no funding, i.e. $\Delta T_{\mathrm{A}} = 0$, we recover the classical model with $C_{\mu} = C_{\mathrm{A}}$.

Instead of a single (zero-bond like) funding, we can also distribute the time $T_{i}$ cost $C_{\mu}(T_{i})$ to a collection of $m_{i}$ equal value loans with repayments in $T^{k}_{i}$, $k=i+1, \ldots i+m_{i}$. In this case the $T_{j}$ cost become
\begin{equation}
    \label{eq:dice:fundingOfAbatementCost2}
    C_{\mathrm{A}}(T_{j}) \ := \sum_{i \ : \ \exists \ T^{k}_{i} = T_{j}} \ \frac{1}{m_{i}} \ C_{\mu}(T_{i}) \ \left( 1 + FR(T_{i}, T_{j};T_{i}) \ (T_{j}-T_{i}) \right) \text{.}
\end{equation}

From a valuation (discounting) perspective, both funding approaches have no effect, as the accrual is compensated by discounting. However, the change in timing might introduce an effect in the DICE model due to the way abatement and damage costs are associated and temporal differences in each generation's utility or capacity.

For a fixed emission pathway, the modification leaves present-value abatement costs unchanged (accrual cancels with discounting). The equity effect comes from changes in timing, not value creation.

Since damages occur instantaneously, we do not consider funding of these. The total cost is given by $C(t) := C_{\mathrm{A}}(t) + C_{\mathrm{D}}(t)$.

In our numerical experiments, we distributed the funding according to \eqref{eq:dice:fundingOfAbatementCost2} over 60 years ($m_{i} = 60$).

\subsubsection{Non-Linear Financing Costs}
\label{sec:nonlinear_discounting}

The present value of an unsecured financial cash-flow is defined by a discount factor times the cash-flow. If the future cash-flow is subject to default, the discount factor is lower, reflecting an additional value reduction due to the risk of (partial) default. As default is not an option for future damage costs, it appears as if the risk-free discount factor should apply. However, since no hedging strategy exists, a risk free funding is not possible. Thus, additional costs may be incurred to secure the unsecured funding \cite{fries_non-linear_2021}. Since damage cost may become very large - much larger than funds provided by standard financial markets - it is reasonable to assume that these additional funding cost become (over-proportionally) large for larger damages. We model this by optionally adding non-linear financing cost using a non-linear funding model \cite{fries_non-linear_2021}. 

This model allows the discount factor to depend on the magnitude of the cash-flow. Thus the funding of larger cash-flows requires a premium to compensate for a larger default risk or other frictions. For our application, this means that larger damages get a larger weight. Consequently, the abatement policy optimization will prefer emissions paths that avoid very large damages, which is beneficial with respect to intergenerational equity.

We modify the damage cost function. Let $C^{\circ}_{\mathrm{D}}(t)$ denote the damage cost of the classical model, i.e.~the quantity that was formerly denoted by $C_{\mathrm{D}}(t)$. We then redefine the effective damage costs as
\begin{equation*}
    C_{\mathrm{D}}(t) \ = \  C^{\circ}_{\mathrm{D}}(t) \cdot DC(C^{\circ}_{\mathrm{D}}(t); t) \text{,}
\end{equation*}
where $DC(C^{\circ}_{\mathrm{D}}(t); t)$ is the \emph{default compensation factor} modeling over-proportional cost of funding large projects, \cite{fries_non-linear_2021}. The factor $DC$ is equal to 1 for modest damage values and larger for large damage values.

As $C^{\circ}_{\mathrm{D}}(t)$ represents a time-value, it is natural that the default compensation factor depends on a normalized value only, e.g., percentage of the GDP. In our numerical experiment we used $DC(C^{\circ}_{\mathrm{D}}(t); t) \ := \ DC^{*}( \frac{C^{\circ}_{\mathrm{D}}(t)}{GDP(t)} )$ with $DC^{*}(x) = 10 \%$ for $x \geq 3 \%$ and $DC^{*}(x) = 1$ otherwise, basically limiting costs to 3 \% of GDP.

As the abatement cost are usually smaller and part of a more planned process, we do not consider a default compensation factor for the abatement cost.

\begin{comment}
    A point we should make is the difference between abatement cost and damage cost.
    The abatement cost ooccure in time $T1$. The associated damage occurs in time $T2$.
    In the classical model the interest rates will establish a weight between the cost.
    Having stochastic interest rates, this weight becomes stochastic.

    In expectation the model will see not difference in the decision to the model with deterministic interest rates.

    The time-lag between abatement and damage will matter, if we consider an objective function different from expectation, e.g., an expected shortfall.
\end{comment}

\subsection{Implementation Details}

The numerical experiments were conducted with an annual time-discretization $\Delta t_{i} = 1$, using a Monte-Carlo simulation with 10000 sample paths. Sensitivities were obtained through adjoint automatic differentiation. Optimization was performed with an ADAM optimizer.

\clearpage

%
% References
%

%TC:ignore

{
\bibliographystyle{naturemag-custom}
\bibliography{references}

\begin{thebibliography}{10}
\expandafter\ifx\csname url\endcsname\relax
  \def\url#1{\texttt{#1}}\fi
\expandafter\ifx\csname urlprefix\endcsname\relax\def\urlprefix{URL }\fi
\providecommand{\bibinfo}[2]{#2}
\providecommand{\eprint}[2][]{\url{#2}}

\bibitem{masson-delmotte_climate_2021}
\bibinfo{editor}{Masson-Delmotte, V.} \emph{et~al.} (eds.)
  \emph{\bibinfo{title}{Climate {Change} 2021: {The} {Physical} {Science}
  {Basis}. {Contribution} of {Working} {Group} {I} to the {Sixth} {Assessment}
  {Report} of the {Intergovernmental} {Panel} on {Climate} {Change}}}.
\newblock {IPCC} (\bibinfo{publisher}{Cambridge University Press},
  \bibinfo{year}{2021}).

\bibitem{portner_ipcc_2022}
\bibinfo{editor}{Pörtner, H.-O.} \emph{et~al.} (eds.)
  \emph{\bibinfo{title}{{IPCC}, 2022: {Climate} {Change} 2022: {Impacts},
  {Adaptation}, and {Vulnerability}. {Contribution} of {Working} {Group} {II}
  to the {Sixth} {Assessment} {Report} of the {Intergovernmental} {Panel} on
  {Climate} {Change}. {Cambridge} {University} {Press}.}}
  (\bibinfo{publisher}{Cambridge University Press}, \bibinfo{year}{2022}).

\bibitem{solomon_persistence_2010}
\bibinfo{author}{Solomon, S.} \emph{et~al.}
\newblock \bibinfo{title}{Persistence of climate changes due to a range of
  greenhouse gases}.
\newblock \emph{\bibinfo{journal}{Proceedings of the National Academy of
  Sciences}} \textbf{\bibinfo{volume}{107}}, \bibinfo{pages}{18354--18359}
  (\bibinfo{year}{2010}).
\newblock DOI
  \href{https://doi.org/10.1073/pnas.1006282107}{10.1073/pnas.1006282107}.

\bibitem{weyant_contributions_2017}
\bibinfo{author}{Weyant, J.}
\newblock \bibinfo{title}{Some {Contributions} of {Integrated} {Assessment}
  {Models} of {Global} {Climate} {Change}}.
\newblock \emph{\bibinfo{journal}{Review of Environmental Economics and
  Policy}} \textbf{\bibinfo{volume}{11}}, \bibinfo{pages}{115--137}
  (\bibinfo{year}{2017}).
\newblock DOI \href{https://doi.org/10.1093/reep/rew018}{10.1093/reep/rew018}.

\bibitem{nordhaus_evolution_2018}
\bibinfo{author}{Nordhaus, W.}
\newblock \bibinfo{title}{Evolution of modeling of the economics of global
  warming: changes in the {DICE} model, 1992–2017}.
\newblock \emph{\bibinfo{journal}{Climatic Change}}
  \textbf{\bibinfo{volume}{148}}, \bibinfo{pages}{623--640}
  (\bibinfo{year}{2018}).
\newblock DOI
  \href{https://doi.org/10.1007/s10584-018-2218-y}{10.1007/s10584-018-2218-y}.

\bibitem{Nordhaus2013}
\bibinfo{author}{Nordhaus, W.}
\newblock \bibinfo{title}{The climate casino: {Risk}, uncertainty, and
  economics for a warming world}.
\newblock \emph{\bibinfo{journal}{The Climate Casino: Risk, Uncertainty, and
  Economics for a Warming World}} \bibinfo{pages}{1--378}
  (\bibinfo{year}{2013}).
\newblock DOI
  \href{https://doi.org/10.1080/14697688.2014.887853}{10.1080/14697688.2014.887853}.

\bibitem{Nordhaus2017}
\bibinfo{author}{Nordhaus, W.~D.}
\newblock \bibinfo{title}{Revisiting the social cost of carbon}.
\newblock \emph{\bibinfo{journal}{Proceedings of the National Academy of
  Sciences of the United States of America}} \textbf{\bibinfo{volume}{114}},
  \bibinfo{pages}{1518--1523} (\bibinfo{year}{2017}).
\newblock DOI
  \href{https://doi.org/10.1073/pnas.1609244114}{10.1073/pnas.1609244114}.

\bibitem{barrage_policies_2024}
\bibinfo{author}{Barrage, L.} \& \bibinfo{author}{Nordhaus, W.}
\newblock \bibinfo{title}{Policies, projections, and the social cost of carbon:
  {Results} from the {DICE}-2023 model}.
\newblock \emph{\bibinfo{journal}{Proceedings of the National Academy of
  Sciences}} \textbf{\bibinfo{volume}{121}}, \bibinfo{pages}{e2312030121}
  (\bibinfo{year}{2024}).
\newblock DOI
  \href{https://doi.org/10.1073/pnas.2312030121}{10.1073/pnas.2312030121}.

\bibitem{faulwasser_towards_2018}
\bibinfo{author}{Faulwasser, T.}, \bibinfo{author}{Nydestedt, R.},
  \bibinfo{author}{Kellett, C.~M.} \& \bibinfo{author}{Weller, S.~R.}
\newblock \bibinfo{title}{Towards a {FAIR}-{DICE} {IAM}: {Combining} {DICE} and
  {FAIR} {Models}}.
\newblock \emph{\bibinfo{journal}{IFAC-PapersOnLine}}
  \textbf{\bibinfo{volume}{51}}, \bibinfo{pages}{126--131}
  (\bibinfo{year}{2018}).
\newblock DOI
  \href{https://doi.org/10.1016/j.ifacol.2018.06.222}{10.1016/j.ifacol.2018.06.222}.

\bibitem{wirths_permafrost_2018}
\bibinfo{author}{Wirths, H.}, \bibinfo{author}{Rathmann, J.} \&
  \bibinfo{author}{Michaelis, P.}
\newblock \bibinfo{title}{The permafrost carbon feedback in {DICE}-{2013R}
  modeling and empirical results}.
\newblock \emph{\bibinfo{journal}{Environmental Economics and Policy Studies}}
  \textbf{\bibinfo{volume}{20}}, \bibinfo{pages}{109--124}
  (\bibinfo{year}{2018}).
\newblock DOI
  \href{https://doi.org/10.1007/s10018-017-0186-5}{10.1007/s10018-017-0186-5}.

\bibitem{Grubb2021ModelingMO}
\bibinfo{author}{Grubb, M.}, \bibinfo{author}{Wieners, C.~E.} \&
  \bibinfo{author}{Yang, P.}
\newblock \bibinfo{title}{Modeling myths: On dice and dynamic realism in
  integrated assessment models of climate change mitigation}.
\newblock \emph{\bibinfo{journal}{Wiley Interdisciplinary Reviews: Climate
  Change}} \textbf{\bibinfo{volume}{12}} (\bibinfo{year}{2021}).

\bibitem{Glanemann2020}
\bibinfo{author}{Glanemann, N.}, \bibinfo{author}{Willner, S.~N.} \&
  \bibinfo{author}{Levermann, A.}
\newblock \bibinfo{title}{Paris {Climate} {Agreement} passes the cost-benefit
  test}.
\newblock \emph{\bibinfo{journal}{Nature Communications}}
  \textbf{\bibinfo{volume}{11}}, \bibinfo{pages}{1--11} (\bibinfo{year}{2020}).
\newblock DOI
  \href{https://doi.org/10.1038/s41467-019-13961-1}{10.1038/s41467-019-13961-1}.

\bibitem{Hansel2020}
\bibinfo{author}{Hänsel, M.~C.} \emph{et~al.}
\newblock \bibinfo{title}{Climate economics support for the {UN} climate
  targets}.
\newblock \emph{\bibinfo{journal}{Nature Climate Change}}
  (\bibinfo{year}{2020}).
\newblock DOI
  \href{https://doi.org/10.1038/s41558-020-0833-x}{10.1038/s41558-020-0833-x}.

\bibitem{crost_optimal_2013}
\bibinfo{author}{Crost, B.} \& \bibinfo{author}{Traeger, C.~P.}
\newblock \bibinfo{title}{Optimal climate policy: {Uncertainty} versus {Monte}
  {Carlo}}.
\newblock \emph{\bibinfo{journal}{Economics Letters}}
  \textbf{\bibinfo{volume}{120}}, \bibinfo{pages}{552--558}
  (\bibinfo{year}{2013}).
\newblock DOI
  \href{https://doi.org/10.1016/j.econlet.2013.05.019}{10.1016/j.econlet.2013.05.019}.

\bibitem{lemoine_watch_2014}
\bibinfo{author}{Lemoine, D.} \& \bibinfo{author}{Traeger, C.}
\newblock \bibinfo{title}{Watch {Your} {Step}: {Optimal} {Policy} in a
  {Tipping} {Climate}}.
\newblock \emph{\bibinfo{journal}{American Economic Journal: Economic Policy}}
  \textbf{\bibinfo{volume}{6}}, \bibinfo{pages}{137--166}
  (\bibinfo{year}{2014}).
\newblock DOI \href{https://doi.org/10.1257/pol.6.1.137}{10.1257/pol.6.1.137}.

\bibitem{jensen_optimal_2014}
\bibinfo{author}{Jensen, S.} \& \bibinfo{author}{Traeger, C.~P.}
\newblock \bibinfo{title}{Optimal climate change mitigation under long-term
  growth uncertainty: {Stochastic} integrated assessment and analytic
  findings}.
\newblock \emph{\bibinfo{journal}{European Economic Review}}
  \textbf{\bibinfo{volume}{69}}, \bibinfo{pages}{104--125}
  (\bibinfo{year}{2014}).
\newblock DOI
  \href{https://doi.org/10.1016/j.euroecorev.2014.01.008}{10.1016/j.euroecorev.2014.01.008}.

\bibitem{cai_environmental_2015}
\bibinfo{author}{Cai, Y.}, \bibinfo{author}{Judd, K.~L.},
  \bibinfo{author}{Lenton, T.~M.}, \bibinfo{author}{Lontzek, T.~S.} \&
  \bibinfo{author}{Narita, D.}
\newblock \bibinfo{title}{Environmental tipping points significantly affect the
  cost-benefit assessment of climate policies}.
\newblock \emph{\bibinfo{journal}{Proceedings of the National Academy of
  Sciences}} \textbf{\bibinfo{volume}{112}}, \bibinfo{pages}{4606--4611}
  (\bibinfo{year}{2015}).
\newblock DOI
  \href{https://doi.org/10.1073/pnas.1503890112}{10.1073/pnas.1503890112}.

\bibitem{cai_risk_2016}
\bibinfo{author}{Cai, Y.}, \bibinfo{author}{Lenton, T.~M.} \&
  \bibinfo{author}{Lontzek, T.~S.}
\newblock \bibinfo{title}{Risk of multiple interacting tipping points should
  encourage rapid {CO2} emission reduction}.
\newblock \emph{\bibinfo{journal}{Nature Climate Change}}
  \textbf{\bibinfo{volume}{6}}, \bibinfo{pages}{520--525}
  (\bibinfo{year}{2016}).
\newblock DOI
  \href{https://doi.org/10.1038/nclimate2964}{10.1038/nclimate2964}.

\bibitem{lemoine_economics_2016}
\bibinfo{author}{Lemoine, D.} \& \bibinfo{author}{Traeger, C.~P.}
\newblock \bibinfo{title}{Economics of tipping the climate dominoes}.
\newblock \emph{\bibinfo{journal}{Nature Climate Change}}
  \textbf{\bibinfo{volume}{6}}, \bibinfo{pages}{514--519}
  (\bibinfo{year}{2016}).
\newblock DOI
  \href{https://doi.org/10.1038/nclimate2902}{10.1038/nclimate2902}.

\bibitem{dietz_economic_2021}
\bibinfo{author}{Dietz, S.}, \bibinfo{author}{Rising, J.},
  \bibinfo{author}{Stoerk, T.} \& \bibinfo{author}{Wagner, G.}
\newblock \bibinfo{title}{Economic impacts of tipping points in the climate
  system}.
\newblock \emph{\bibinfo{journal}{Proceedings of the National Academy of
  Sciences}} \textbf{\bibinfo{volume}{118}}, \bibinfo{pages}{e2103081118}
  (\bibinfo{year}{2021}).
\newblock DOI
  \href{https://doi.org/10.1073/pnas.2103081118}{10.1073/pnas.2103081118}.

\bibitem{Ackerman2010a}
\bibinfo{author}{Ackerman, F.}, \bibinfo{author}{Stanton, E.~A.} \&
  \bibinfo{author}{Bueno, R.}
\newblock \bibinfo{title}{Fat tails, exponents, extreme uncertainty:
  {Simulating} catastrophe in {DICE}}.
\newblock \emph{\bibinfo{journal}{Ecological Economics}}
  \textbf{\bibinfo{volume}{69}}, \bibinfo{pages}{1657--1665}
  (\bibinfo{year}{2010}).
\newblock DOI
  \href{https://doi.org/10.1016/j.ecolecon.2010.03.013}{10.1016/j.ecolecon.2010.03.013}.

\bibitem{ikefuji_expected_2020}
\bibinfo{author}{Ikefuji, M.}, \bibinfo{author}{Laeven, R. J.~A.},
  \bibinfo{author}{Magnus, J.~R.} \& \bibinfo{author}{Muris, C.}
\newblock \bibinfo{title}{Expected utility and catastrophic risk in a
  stochastic economy–climate model}.
\newblock \emph{\bibinfo{journal}{Journal of Econometrics}}
  \textbf{\bibinfo{volume}{214}}, \bibinfo{pages}{110--129}
  (\bibinfo{year}{2020}).
\newblock DOI
  \href{https://doi.org/10.1016/j.jeconom.2019.05.007}{10.1016/j.jeconom.2019.05.007}.

\bibitem{Hepburn2007}
\bibinfo{author}{Hepburn, C.} \& \bibinfo{author}{Beckerman, W.}
\newblock \bibinfo{title}{Ethics of the discount rate in the {Stern} {Review}
  on the economics of climate change}.
\newblock \emph{\bibinfo{journal}{World Economics}}
  \textbf{\bibinfo{volume}{8}}, \bibinfo{pages}{187--211}
  (\bibinfo{year}{2007}).

\bibitem{Gollier2010}
\bibinfo{author}{Gollier, C.} \& \bibinfo{author}{Weitzman, M.~L.}
\newblock \bibinfo{title}{How should the distant future be discounted when
  discount rates are uncertain?}
\newblock \emph{\bibinfo{journal}{Economics Letters}}
  \textbf{\bibinfo{volume}{107}}, \bibinfo{pages}{350--353}
  (\bibinfo{year}{2010}).
\newblock DOI
  \href{https://doi.org/10.1016/j.econlet.2010.03.001}{10.1016/j.econlet.2010.03.001}.

\bibitem{van_der_ploeg_simple_2019}
\bibinfo{author}{van~der Ploeg, F.} \& \bibinfo{author}{Rezai, A.}
\newblock \bibinfo{title}{Simple {Rules} for {Climate} {Policy} and
  {Integrated} {Assessment}}.
\newblock \emph{\bibinfo{journal}{Environmental and Resource Economics}}
  \textbf{\bibinfo{volume}{72}}, \bibinfo{pages}{77--108}
  (\bibinfo{year}{2019}).
\newblock DOI
  \href{https://doi.org/10.1007/s10640-018-0280-6}{10.1007/s10640-018-0280-6}.

\bibitem{VanDerPloeg2020}
\bibinfo{author}{Van Der~Ploeg, F.}
\newblock \bibinfo{title}{Discounting and {Climate} {Policy}}.
\newblock \emph{\bibinfo{journal}{CESifo Working Papers}}
  (\bibinfo{year}{2020}).
\newblock DOI
  \href{https://doi.org/10.2139/ssrn.3657977}{10.2139/ssrn.3657977}.

\bibitem{wagner_carbon_2020}
\bibinfo{author}{Wagner, G.} \& \bibinfo{author}{Hoggan, W.~H.}
\newblock \bibinfo{title}{Carbon {Prices}, {Preferences}, and the {Timing} of
  {Uncertainty}} (\bibinfo{year}{2020}).

\bibitem{adler_priority_2017}
\bibinfo{author}{Adler, M.} \emph{et~al.}
\newblock \bibinfo{title}{Priority for the worse-off and the social cost of
  carbon}.
\newblock \emph{\bibinfo{journal}{Nature Climate Change}}
  \textbf{\bibinfo{volume}{7}}, \bibinfo{pages}{443--449}
  (\bibinfo{year}{2017}).
\newblock DOI
  \href{https://doi.org/10.1038/nclimate3298}{10.1038/nclimate3298}.

\bibitem{padilla_intergenerational_2002}
\bibinfo{author}{Padilla, E.}
\newblock \bibinfo{title}{Intergenerational equity and sustainability}.
\newblock \emph{\bibinfo{journal}{Ecological Economics}}
  \textbf{\bibinfo{volume}{41}}, \bibinfo{pages}{69--83}
  (\bibinfo{year}{2002}).
\newblock DOI
  \href{https://doi.org/10.1016/S0921-8009(02)00026-5}{10.1016/S0921-8009(02)00026-5}.

\bibitem{emmerling_discounting_2018}
\bibinfo{author}{Emmerling, J.}
\newblock \bibinfo{title}{Discounting and intragenerational equity}.
\newblock \emph{\bibinfo{journal}{Environment and Development Economics}}
  \textbf{\bibinfo{volume}{23}}, \bibinfo{pages}{19--36}
  (\bibinfo{year}{2018}).
\newblock DOI
  \href{https://doi.org/10.1017/S1355770X17000365}{10.1017/S1355770X17000365}.

\bibitem{bauer_rising_2021}
\bibinfo{author}{Bauer, M.~D.} \& \bibinfo{author}{Rudebusch, G.~D.}
\newblock \bibinfo{title}{The {Rising} {Cost} of {Climate} {Change}: {Evidence}
  from the {Bond} {Market}}.
\newblock \emph{\bibinfo{journal}{The Review of Economics and Statistics}}
  \bibinfo{pages}{1--45} (\bibinfo{year}{2021}).
\newblock DOI
  \href{https://doi.org/10.1162/rest\_a\_01109}{10.1162/rest\_a\_01109}.

\bibitem{newell_discounting_2022}
\bibinfo{author}{Newell, R.~G.}, \bibinfo{author}{Pizer, W.~A.} \&
  \bibinfo{author}{Prest, B.~C.}
\newblock \bibinfo{title}{A {Discounting} {Rule} for the {Social} {Cost} of
  {Carbon}}.
\newblock \emph{\bibinfo{journal}{Journal of the Association of Environmental
  and Resource Economists}} \textbf{\bibinfo{volume}{9}},
  \bibinfo{pages}{1017--1046} (\bibinfo{year}{2022}).
\newblock DOI \href{https://doi.org/10.1086/718145}{10.1086/718145}.

\bibitem{ANDERSEN2020102290}
\bibinfo{author}{Andersen, T.~M.}, \bibinfo{author}{Bhattacharya, J.} \&
  \bibinfo{author}{Liu, P.}
\newblock \bibinfo{title}{Resolving intergenerational conflict over the
  environment under the pareto criterion}.
\newblock \emph{\bibinfo{journal}{Journal of Environmental Economics and
  Management}} \textbf{\bibinfo{volume}{100}}, \bibinfo{pages}{102290}
  (\bibinfo{year}{2020}).
\newblock DOI
  \href{https://doi.org/https://doi.org/10.1016/j.jeem.2019.102290}{https://doi.org/10.1016/j.jeem.2019.102290}.

\bibitem{Kotlikoff2021CarbonTaxation}
\bibinfo{author}{Kotlikoff, L.}, \bibinfo{author}{Kubler, F.},
  \bibinfo{author}{Polbin, A.}, \bibinfo{author}{Sachs, J.} \&
  \bibinfo{author}{Scheidegger, S.}
\newblock \bibinfo{title}{Making carbon taxation a generational win win}.
\newblock \emph{\bibinfo{journal}{International Economic Review}}
  \textbf{\bibinfo{volume}{62}}, \bibinfo{pages}{3--46} (\bibinfo{year}{2021}).
\newblock DOI
  \href{https://doi.org/https://doi.org/10.1111/iere.12483}{https://doi.org/10.1111/iere.12483}.

\bibitem{daniel_pushing_2013}
\bibinfo{author}{Daniel, B.~C.} \& \bibinfo{author}{Shiamptanis, C.}
\newblock \bibinfo{title}{Pushing the limit? {Fiscal} policy in the {European}
  {Monetary} {Union}}.
\newblock \emph{\bibinfo{journal}{Journal of Economic Dynamics and Control}}
  \textbf{\bibinfo{volume}{37}}, \bibinfo{pages}{2307--2321}
  (\bibinfo{year}{2013}).
\newblock DOI
  \href{https://doi.org/10.1016/j.jedc.2013.06.003}{10.1016/j.jedc.2013.06.003}.

\bibitem{finmathlib}
\bibinfo{author}{Fries, C.~P.}
\newblock \bibinfo{title}{finmath-lib: Mathematical finance library: Algorithms
  and metholodgies related to mathematical finance}
  (\bibinfo{year}{2012-2025}).
\newblock \urlprefix\url{http://finmath.net/finmath-lib/}.

\bibitem{noauthor_world_nodate}
\bibinfo{title}{World {Population} {Prospects} - {Population} {Division} -
  {United} {Nations}}.
\newblock
  \urlprefix\url{https://population.un.org/wpp/Download/Standard/MostUsed/}.

\bibitem{Fries2024FairShareOfGDP}
\bibinfo{author}{Fries, C.~P.}
\newblock \bibinfo{title}{Fair share of gdp to mitigate climate change cost
  (according to dice)}.
\newblock \emph{\bibinfo{journal}{SSRN Electronic Journal}}
  (\bibinfo{year}{2024}).
\newblock DOI
  \href{https://doi.org/10.2139/ssrn.5040331}{10.2139/ssrn.5040331}.

\bibitem{Burke2015}
\bibinfo{author}{Burke, M.}, \bibinfo{author}{Hsiang, S.~M.} \&
  \bibinfo{author}{Miguel, E.}
\newblock \bibinfo{title}{Global non-linear effect of temperature on economic
  production}.
\newblock \emph{\bibinfo{journal}{Nature}} \textbf{\bibinfo{volume}{527}},
  \bibinfo{pages}{235--239} (\bibinfo{year}{2015}).
\newblock DOI \href{https://doi.org/10.1038/nature15725}{10.1038/nature15725}.

\bibitem{kotz_day--day_2021}
\bibinfo{author}{Kotz, M.}, \bibinfo{author}{Wenz, L.},
  \bibinfo{author}{Stechemesser, A.}, \bibinfo{author}{Kalkuhl, M.} \&
  \bibinfo{author}{Levermann, A.}
\newblock \bibinfo{title}{Day-to-day temperature variability reduces economic
  growth}.
\newblock \emph{\bibinfo{journal}{Nature Climate Change}}
  \textbf{\bibinfo{volume}{11}}, \bibinfo{pages}{319--325}
  (\bibinfo{year}{2021}).
\newblock DOI
  \href{https://doi.org/10.1038/s41558-020-00985-5}{10.1038/s41558-020-00985-5}.

\bibitem{kotz_effect_2022}
\bibinfo{author}{Kotz, M.}, \bibinfo{author}{Levermann, A.} \&
  \bibinfo{author}{Wenz, L.}
\newblock \bibinfo{title}{The effect of rainfall changes on economic
  production}.
\newblock \emph{\bibinfo{journal}{Nature}} \textbf{\bibinfo{volume}{601}},
  \bibinfo{pages}{223--227} (\bibinfo{year}{2022}).
\newblock DOI
  \href{https://doi.org/10.1038/s41586-021-04283-8}{10.1038/s41586-021-04283-8}.

\bibitem{andreoni_inequality_2023}
\bibinfo{author}{Andreoni, P.}, \bibinfo{author}{Emmerling, J.} \&
  \bibinfo{author}{Tavoni, M.}
\newblock \bibinfo{title}{Inequality repercussions of financing negative
  emissions}.
\newblock \emph{\bibinfo{journal}{Nature Climate Change}} \bibinfo{pages}{1--7}
  (\bibinfo{year}{2023}).
\newblock DOI
  \href{https://doi.org/10.1038/s41558-023-01870-7}{10.1038/s41558-023-01870-7}.

\bibitem{rennert_comprehensive_2022}
\bibinfo{author}{Rennert, K.} \emph{et~al.}
\newblock \bibinfo{title}{Comprehensive {Evidence} {Implies} a {Higher}
  {Social} {Cost} of {CO2}}.
\newblock \emph{\bibinfo{journal}{Nature}} \bibinfo{pages}{1--3}
  (\bibinfo{year}{2022}).
\newblock DOI
  \href{https://doi.org/10.1038/s41586-022-05224-9}{10.1038/s41586-022-05224-9}.

\bibitem{van_der_wijst_new_2023}
\bibinfo{author}{van~der Wijst, K.-I.} \emph{et~al.}
\newblock \bibinfo{title}{New damage curves and multimodel analysis suggest
  lower optimal temperature}.
\newblock \emph{\bibinfo{journal}{Nature Climate Change}} \bibinfo{pages}{1--8}
  (\bibinfo{year}{2023}).
\newblock DOI
  \href{https://doi.org/10.1038/s41558-023-01636-1}{10.1038/s41558-023-01636-1}.

\bibitem{Fries2023CO2}
\bibinfo{author}{Fries, C.~P.}
\newblock \bibinfo{title}{Implied co2-price, carbon interest rate, fair share
  of gdp, and the extension of an integrated assessment model with a climate
  transformation fund}.
\newblock \emph{\bibinfo{journal}{SSRN Electronic Journal}}
  (\bibinfo{year}{2023}).
\newblock DOI
  \href{https://doi.org/10.2139/ssrn.3284722}{10.2139/ssrn.3284722}.

\bibitem{fries_intergenerational_2023}
\bibinfo{author}{Fries, C.~P.} \& \bibinfo{author}{Quante, L.}
\newblock \bibinfo{title}{Intergenerational {Equity} in {Models} of {Climate}
  {Change} {Mitigation}: {Stochastic} {Interest} {Rates} introduce {Adverse}
  {Effects}, but ({Non}-linear) {Funding} {Costs} can {Improve}
  {Intergenerational} {Equity}} (\bibinfo{year}{2023}).
\newblock \urlprefix\url{https://papers.ssrn.com/abstract=4005846}.
\newblock DOI
  \href{https://doi.org/10.2139/ssrn.4005846}{10.2139/ssrn.4005846}.

\bibitem{fries_non-linear_2021}
\bibinfo{author}{Fries, C.~P.}
\newblock \bibinfo{title}{Non-{Linear} {Discounting} and {Default}
  {Compensation}: {Valuation} of {Non}-{Replicable} {Value} and {Damage}:
  {When} the {Social} {Discount} {Rate} may become {Negative}}
  (\bibinfo{year}{2021}).
\newblock \urlprefix\url{http://arxiv.org/abs/2007.06465}.
\newblock \bibinfo{note}{ArXiv:2007.06465 [q-fin]}.
\newblock DOI
  \href{https://doi.org/10.48550/arXiv.2007.06465}{10.48550/arXiv.2007.06465}.

\bibitem{hull1990pricing}
\bibinfo{author}{Hull, J.} \& \bibinfo{author}{White, A.}
\newblock \bibinfo{title}{Pricing interest-rate-derivative securities}.
\newblock \emph{\bibinfo{journal}{The review of financial studies}}
  \textbf{\bibinfo{volume}{3}}, \bibinfo{pages}{573--592}
  (\bibinfo{year}{1990}).

\bibitem{FriesLectureNotes2007}
\bibinfo{author}{Fries, C.}
\newblock \emph{\bibinfo{title}{Mathematical Finance: Theory, Modeling,
  Implementation}} (\bibinfo{publisher}{Wiley}, \bibinfo{year}{2007}),
  \bibinfo{edition}{hardcover} edn.
\newblock
  \urlprefix\url{https://lead.to/amazon/com/?op=bt&la=en&cu=usd&key=0470047224}.

\bibitem{fries_hull_white2016}
\bibinfo{author}{Fries, C.~P.}
\newblock \bibinfo{title}{A short note on the exact stochastic simulation
  scheme of the hull-white model and its implementation}.
\newblock \emph{\bibinfo{journal}{SSRN Electronic Journal}}
  (\bibinfo{year}{2016}).
\newblock DOI
  \href{https://doi.org/10.2139/ssrn.2737091}{10.2139/ssrn.2737091}.

\bibitem{finmath-climate}
\bibinfo{author}{Fries, C.~P.} \& \bibinfo{author}{Quante, L.}
\newblock \bibinfo{title}{Stochastic and non-linear extensions to integrated
  assessment models} (\bibinfo{year}{2019}).
\newblock \urlprefix\url{https://gitlab.com/finmath/finmath-climate-nonlinear}.

\bibitem{buchholz_discounting_2010}
\bibinfo{author}{Buchholz, W.} \& \bibinfo{author}{Schumacher, J.}
\newblock \bibinfo{title}{Discounting and welfare analysis over time:
  {Choosing} the $\eta$}.
\newblock \emph{\bibinfo{journal}{European Journal of Political Economy}}
  \textbf{\bibinfo{volume}{26}}, \bibinfo{pages}{372--385}
  (\bibinfo{year}{2010}).
\newblock DOI
  \href{https://doi.org/10.1016/j.ejpoleco.2009.11.011}{10.1016/j.ejpoleco.2009.11.011}.

\bibitem{newell_pizer_discounting_2004}
\bibinfo{author}{Newell, R.~G.} \& \bibinfo{author}{Pizer, W.~A.}
\newblock \bibinfo{title}{Uncertain discount rates in climate policy analysis}.
\newblock \emph{\bibinfo{journal}{Energy Policy}}
  \textbf{\bibinfo{volume}{32}}, \bibinfo{pages}{519--529}
  (\bibinfo{year}{2004}).
\newblock DOI
  \href{https://doi.org/https://doi.org/10.1016/S0301-4215(03)00153-8}{https://doi.org/10.1016/S0301-4215(03)00153-8}.

\bibitem{taconet_social_2021}
\bibinfo{author}{Taconet, N.}, \bibinfo{author}{Guivarch, C.} \&
  \bibinfo{author}{Pottier, A.}
\newblock \bibinfo{title}{Social {Cost} of {Carbon} {Under} {Stochastic}
  {Tipping} {Points}}.
\newblock \emph{\bibinfo{journal}{Environmental and Resource Economics}}
  (\bibinfo{year}{2021}).
\newblock DOI
  \href{https://doi.org/10.1007/s10640-021-00549-x}{10.1007/s10640-021-00549-x}.

\bibitem{carleton_valuing_2020}
\bibinfo{author}{Carleton, T.~A.} \emph{et~al.}
\newblock \bibinfo{title}{Valuing the {Global} {Mortality} {Consequences} of
  {Climate} {Change} {Accounting} for {Adaptation} {Costs} and {Benefits}}.
\newblock \bibinfo{type}{Working {Paper}} \bibinfo{number}{27599},
  \bibinfo{institution}{National Bureau of Economic Research}
  (\bibinfo{year}{2020}).
\newblock \urlprefix\url{https://www.nber.org/papers/w27599}.
\newblock \bibinfo{note}{Series: Working Paper Series}.
\newblock DOI \href{https://doi.org/10.3386/w27599}{10.3386/w27599}.

\bibitem{rode_estimating_2021}
\bibinfo{author}{Rode, A.} \emph{et~al.}
\newblock \bibinfo{title}{Estimating a social cost of carbon for global energy
  consumption}.
\newblock \emph{\bibinfo{journal}{Nature}} \textbf{\bibinfo{volume}{598}},
  \bibinfo{pages}{308--314} (\bibinfo{year}{2021}).
\newblock DOI
  \href{https://doi.org/10.1038/s41586-021-03883-8}{10.1038/s41586-021-03883-8}.

\end{thebibliography}
}

\clearpage

\subsection*{Code availability}

The model code for the stochastic DICE model, including the proposed extensions, and the experiments generating the data and figures is available in the following Git repository \url{https://gitlab.com/finmath/finmath-climate-nonlinear}. %with identifier \href{https://doi.org/10.5281/zenodo.8363819}{10.5281/zenodo.8363819}.

%The simulation data that support the findings of this study will be openly available at the public repository for this publication with identifier \todo{\href{https://doi.org/10.5281/zenodo.8363819}{10.5281/zenodo.8363819}}.

\subsection*{Data availability}

The data used in the reported results and figures are available in the Git repository \url{https://gitlab.com/finmath/finmath-climate-nonlinear}.

\subsection*{Supplementary Material}

Supplementary material is available online.

%
% Acknowledgements etc,
%

\subsection*{Acknowledgements}

We thank Max Singhoff for exploring the optimization of the DICE model with the ADAM solver used in our numerical experiments.

L.Q. received funding from the German Federal Ministry of Education and Research (BMBF) under the research projects QUIDIC (01LP1907A) as well as by Deutsche Gesellschaft für Internationale Zusammenarbeit (GIZ) GmbH on behalf of the Government of the Federal Republic of Germany and Federal Ministry for Economic Cooperation and Development (BMZ).

\subsection*{Author Contributions}

C.F. and L.Q. developed the research idea together. C.F. lead the re-implementation of the DICE model and extensions, as well as the experiments with input from L.Q. L.Q. lead the plotting implementation with input from C.F. C.F. and L.Q. analyzed and interpreted results. C.F. derived the analytic expression for generational inequality. C.F and L.Q. wrote the manuscript.

\subsection*{Competing Interests}
The authors declare that they have no competing interests.

\clearpage

%reset counters for SI
\resetlinenumber
\setcounter{page}{1}
\setcounter{figure}{0}
\setcounter{table}{0}

% INCLUDE APPENDIX HERE

%%%%%%%%%%%%%%%%%%%%%%%%%%%%%%%%%%%%%%%%%%%%%%%%%%%%%%%%%%%%%%%%%%%%%%%%%%%%%%%%%%%%%%%%%%%%%%%%%%%

\appendix
\restarttoc

\section*{Supplementary information for}
    \noindent
    \textbf{\Large{\articletitle}}
  {
  
    \noindent Christian P. Fries \orcidlink{0000-0003-4767-2034} and
    Lennart Quante \orcidlink{0000-0003-4942-8254} 
  }

\bigskip
\section*{Contents}
\supplementtoc

\captionsetup{labelfont=bf,list=yes}
\makeatletter
\renewcommand{\fnum@figure}[1]{\figurename\ \thefigure} % remove the : after number
\makeatother
\makeatletter
\renewcommand{\fnum@table}[1]{\tablename\ \thetable} % remove the : after number
\makeatother
\addto\captionsenglish{\renewcommand{\figurename}{Supplementary Figure}}
\renewcommand{\figurename}{Supplementary Figure}
\addto\captionsenglish{\renewcommand{\tablename}{Supplementary Table}}
\renewcommand{\tablename}{Supplementary Table}

\newpage

%%%%%%%%%%%%%%%%%%%%%%%%%%%%%%%%%%%%%%%%%%%%%%%%%%%%%%%%%%%%%%%%%%%%%%%%%%%%%%%%%%%%%%%%%%%%%%%%%%%

\section{Supplementary Figures}

\begin{figure}[hbtp]
    \centering
    \includegraphics[width=\textwidth]{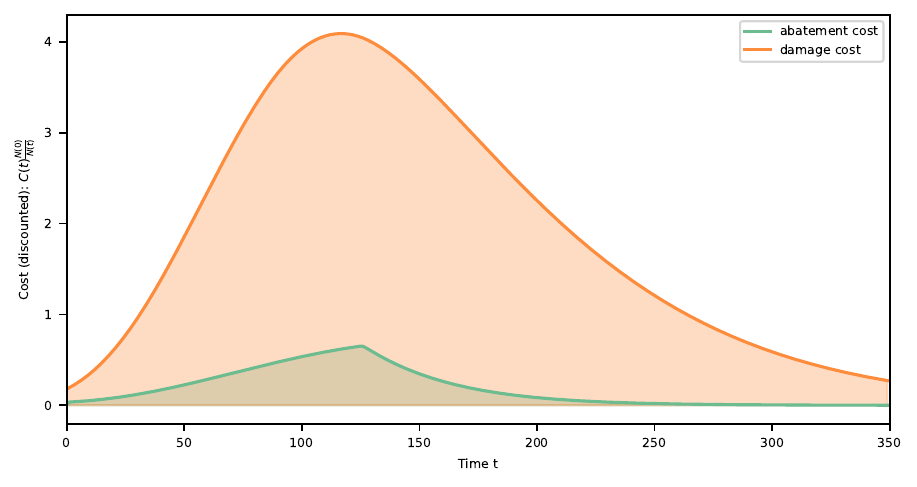}
    \caption[Temporal distribution of cost]{
        The abatement cost $C_{\mathrm{A}}$ (green) and damage cost $C_{\mathrm{D}}$ (red), discounted by $\frac{N(0)}{N(t)}$, in the original DICE model. The optimization is not balancing the temporal distribution of the cost - which is ultimately resulting in an intergenerational inequality.
    }
    \label[supfigure]{fig:cost_abatement_damage_disounted}
\end{figure}

We provide some additional figures. \cref{fig:cost_abatement_damage_disounted} shows the temporal distribution of the cost, which should be compared to the temporal distribution of the cost sensitivity in \cref{fig:abatement_damage_sensitivity_weighted}.

\begin{figure}[hbtp]
    \centering
    \includegraphics[width=\textwidth]{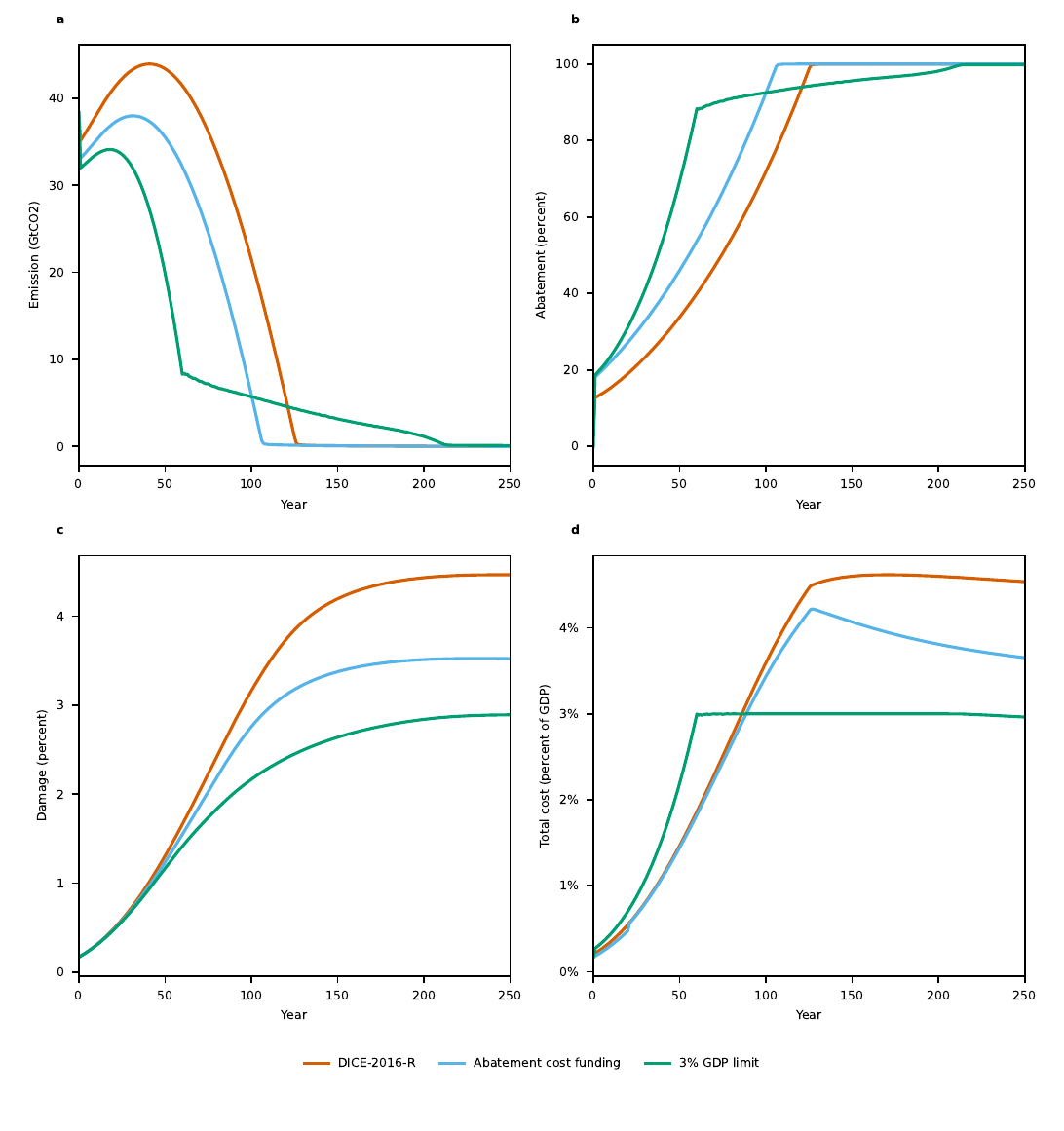}
    \caption[Model quantities under different extensions for the deterministic model]{
		Deterministic model: Financing extensions lead to faster abatement compared to standard model; \textbf{a} emissions, \textbf{b} abatement, \textbf{c} damages, and \textbf{d} total GDP relative cost.
	}
    \label[supfigure]{fig:pathways_deterministic_with_extension}
\end{figure}

\begin{figure}[ht!]
    \centering
    \includegraphics[width=\textwidth]{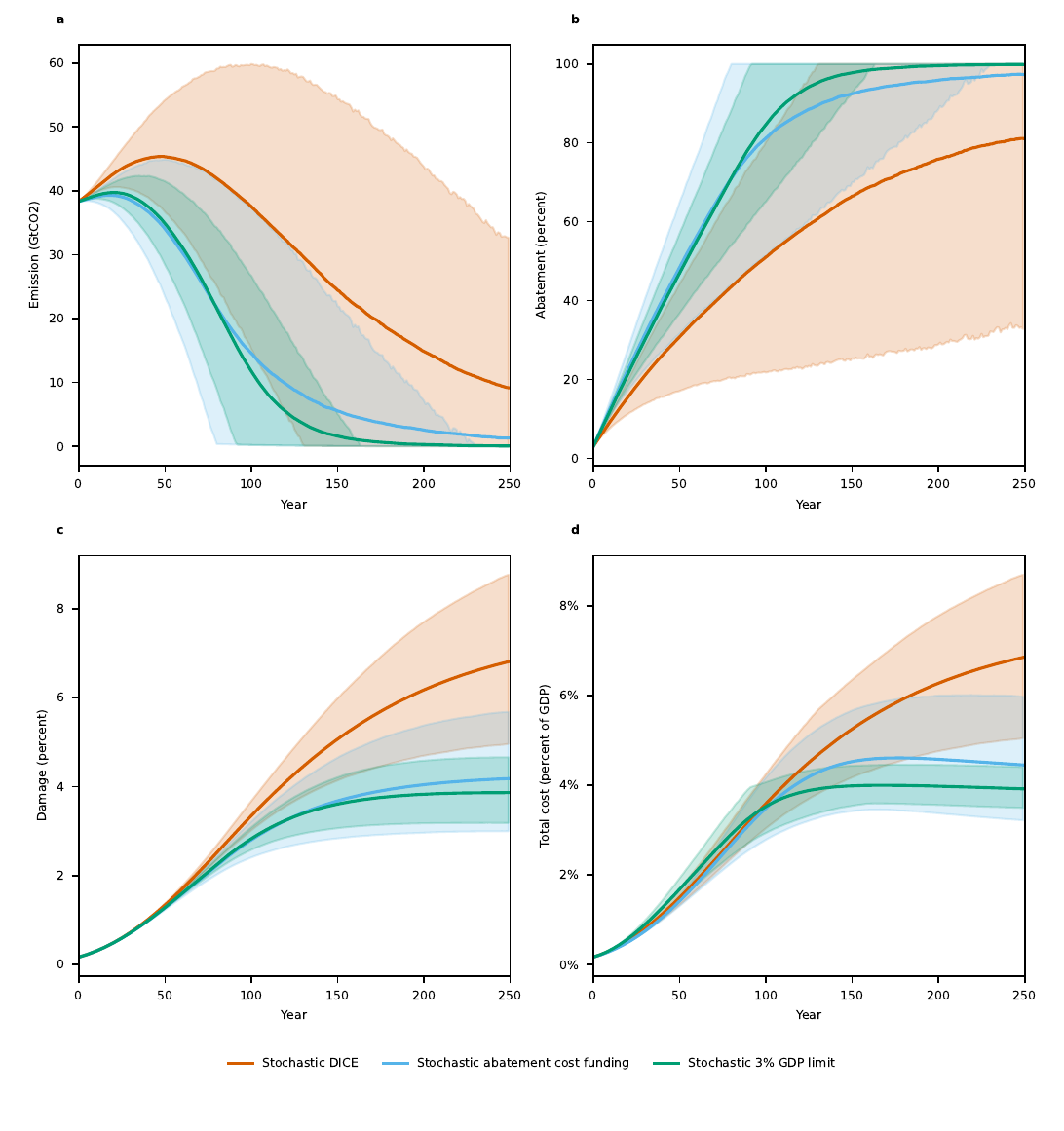}
    \caption[Model quantities under different extensions for the stochastic model]{Stochastic model: Financing extensions lead to faster abatement compared to standard model; \textbf{a} emissions, \textbf{b} abatement, \textbf{c} damages, and \textbf{d} total GDP relative cost. Shaded area shows the 10th to 90th~percentiles (VaR).}
    \label[supfigure]{fig:pathways_stochastic_with_extension}
\end{figure}

\cref{fig:pathways_deterministic_with_extension} and~\cref{fig:pathways_stochastic_with_extension} show the associated model quantities (emission, abatement, damage, and total cost per GDP) for the deterministic and stochastic model, with and without our model extensions.

\begin{figure}[ht!]
    \centering
    \includegraphics[width=\textwidth]{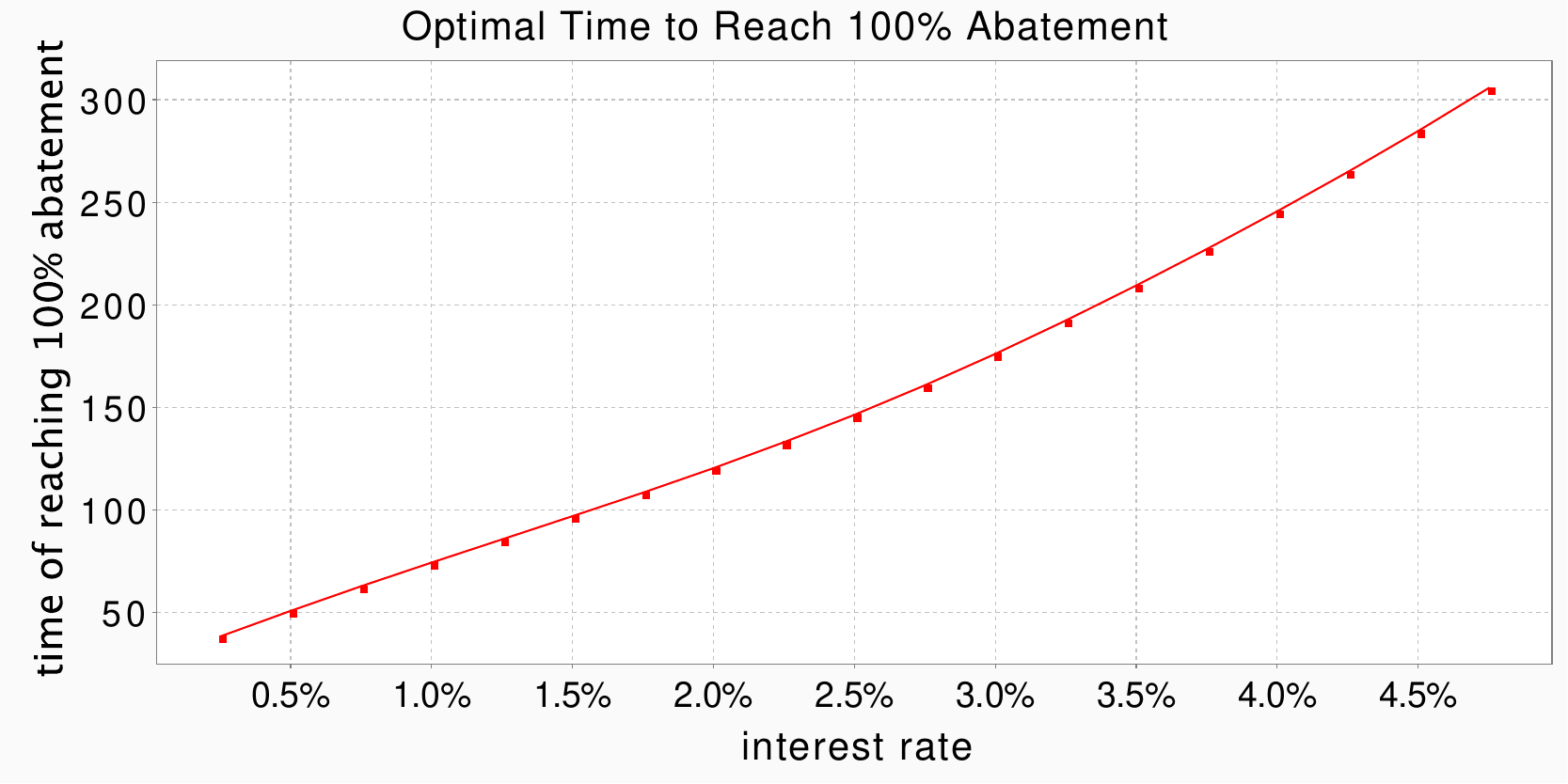}
    \caption[Sensitivity of Abatement to Interest Rate Level]{
        Increasing interest rate level increases the time of 100\% abatement $T^{\mu=1}$.
    }
    \label[supfigure]{fig:opt_abatement_time_by_interest_level}
\end{figure}

\begin{figure}[ht!]
    \centering
    \includegraphics[width=\textwidth]{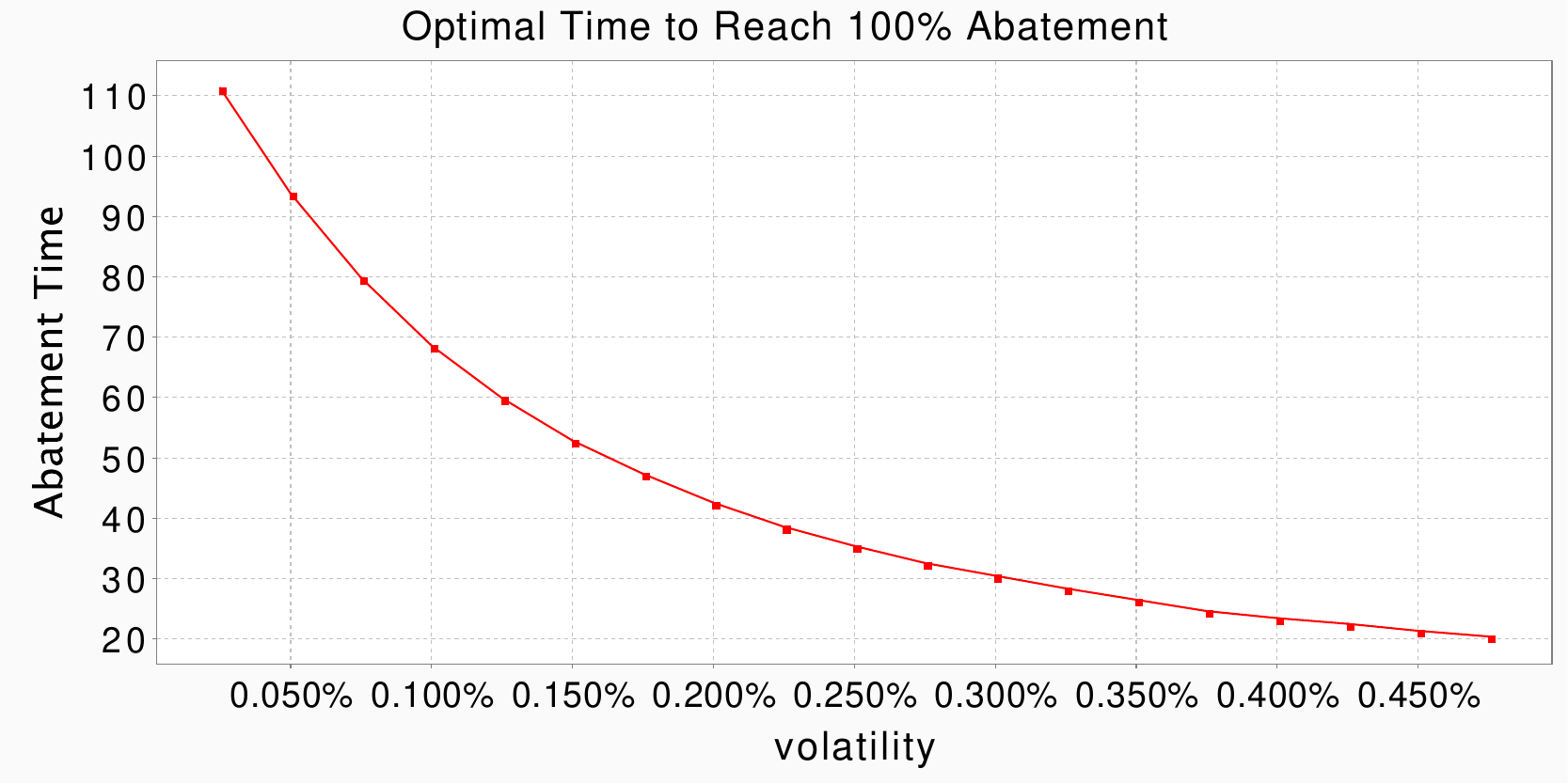}
    \caption[Sensitivity of Abatement to Interest Rate Volatility]{
        Increasing interest rate volatility, i.e. riskiness, decreases the time of 100\% abatement $T^{\mu=1}$.
    }
    \label[supfigure]{fig:opt_abatement_time_by_interest_volatility}
\end{figure}

\cref{fig:opt_abatement_time_by_interest_level} and~\cref{fig:opt_abatement_time_by_interest_volatility} show the sensitivity of the abatement policy on the interest rate level and interest rate volatility.

\begin{footnotesize}
    \listof{figure}{List of Supplementary Figures}
\end{footnotesize}


\begin{figure}[ht!]
    \centering
    \includegraphics[width=\textwidth]{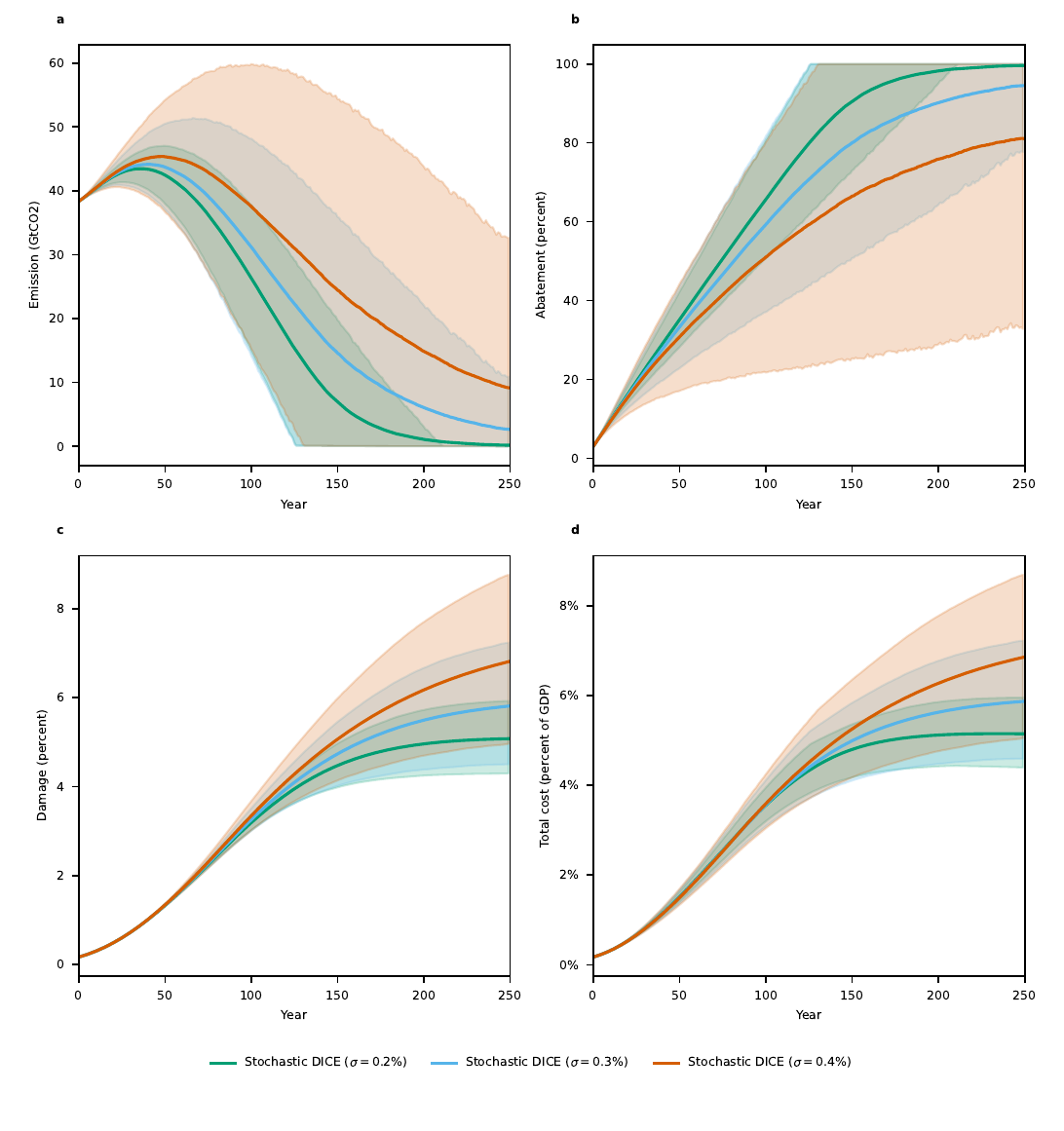}
	\caption[Sensitivity of Emissions, Abatement, Damages and Total Cost to Interest Rate Volatility]{
	    Stochastic DICE Model with Different Interest Rate Volatilities ($\sigma = 0.2 \%$ (green), $\sigma = 0.3 \%$ (blue), $\sigma = 0.4 \%$ (red)). Allowing adaptation of the abatement policy to stochastic interest rates increases intergenerational inequality \emph{and} exhibits a significant risk of even higher intergenerational inequality
	    (\textbf{a}: emissions, \textbf{b}: abatement, \textbf{c}: damages, and \textbf{d}: total GDP relative cost). Shaded area shows the 10th to 90th~percentiles.
	}
    \label[supfigure]{fig:pathways_stochastic_variation}
\end{figure}

\clearpage

%%%%%%%%%%%%%%%%%%%%%%%%%%%%%%%%%%%%%%%%%%%%%%%%%%%%%%%%%%%%%%%%%%%%%%%%%%%%%%%%%%%%%%%%%%%%%%%%%%%

\section{Damage Cost per Abatement Cost upon Abatement Policy Change}
\label{sec:itergeneq:appendix:proof:costdensity}

We show that the \textit{damage cost per abatement cost upon abatement policy change} is a density if $\mu$ is in its equilibrium state.

If $\mu$ denotes the optimal abatement policy we have
\begin{equation}
    \int_{0}^{T} \frac{\mathrm{d}V(t)}{\mathrm{d}C(t)} \frac{\mathrm{d}C_{\mathrm{D}}(t)}{\mathrm{d} \mu(s)} \ \frac{N(0)}{N(t)} \mathrm{d}t
    \ + \
	\int_{0}^{T} \frac{\mathrm{d}V(t)}{\mathrm{d}C(t)} \frac{\mathrm{d}C_{\mathrm{A}}(t)}{\mathrm{d} \mu(s)} \ \frac{N(0)}{N(t)} \mathrm{d}t
	\ \stackrel{!}{=} \ 0 \text{,}
	\tag{\ref{eq:cost_objective_mus}}
\end{equation}
For the time discrete model where $C_{\mathrm{A}}(s)$ depends on $\mu(s)$ only, we find at $s = t_{i}$ that
\begin{equation*}
    \int_{0}^{T} \frac{\mathrm{d}V(t)}{\mathrm{d}C(t)} \frac{\mathrm{d}C_{\mathrm{A}}(t)}{\mathrm{d} \mu(s)} \ \frac{N(0)}{N(t)} \mathrm{d}t \ = \ \frac{\mathrm{d}V(t_{i})}{\mathrm{d}C(t_{i})} \frac{\mathrm{d}C_{\mathrm{A}}(t_{i})}{\mathrm{d} \mu(t_{i})} \ \frac{N(0)}{N(t_{i})} \ \Delta t_{i} \text{.}
\end{equation*}
Hence, for $\Delta t_{i} = 1$ (using $s = t_{i}$)
\begin{equation*}
    \int_{0}^{T} \frac{\mathrm{d}V(t)}{\mathrm{d}C(t)} \frac{\mathrm{d}C_{\mathrm{D}}(t)}{\mathrm{d} \mu(s)} \ \frac{N(0)}{N(t)} \mathrm{d}t
    \ = \ - \frac{\mathrm{d}V(s)}{\mathrm{d}C(s)} \frac{\mathrm{d}C_{\mathrm{A}}(s)}{\mathrm{d} \mu(s)} \ \frac{N(0)}{N(s)} \text{.}
\end{equation*}
and thus
\begin{equation*}
    \int_{0}^{T} - \frac{ \frac{\mathrm{d}V(t)}{\mathrm{d}C(t)} \frac{\mathrm{d}C_{\mathrm{D}}(t)}{\mathrm{d} \mu(s)} \ \frac{N(0)}{N(t)} }{\frac{\mathrm{d}V(s)}{\mathrm{d}C(s)} \frac{\mathrm{d}C_{\mathrm{A}}(s)}{\mathrm{d} \mu(s)} \ \frac{N(0)}{N(s)}} \ \mathrm{d}t
    \ = \ 1 \text{.}
\end{equation*}

\clearpage

%%%%%%%%%%%%%%%%%%%%%%%%%%%%%%%%%%%%%%%%%%%%%%%%%%%%%%%%%%%%%%%%%%%%%%%%%%%%%%%%%%%%%%%%%%%%%%%%%%%

\section{Non-Linear Financing Costs}

We modify the damage cost function as
\begin{equation*}
    C_{\mathrm{D}}(t) \ = \  C^{\circ}_{\mathrm{D}}(t) \cdot DC(C^{\circ}_{\mathrm{D}}(t); t) \text{,}
\end{equation*}
where $C^{\circ}_{\mathrm{D}}(t)$ denote the damage cost of the classical model and $DC(C^{\circ}_{\mathrm{D}}(t); t)$ the \emph{default compensation factor} modeling the over-proportional cost of funding large projects, \cite{fries_non-linear_2021}.

We allow for two different normalizations: with the numéraire $N(t)$ or with the GDP $Y(t)$. Given a function $DC^{*}$, the model for the default compensator is
\begin{equation}
    \label{eq:nonlineardiscounting:relativetoNumeraire}
    DC\left(C^{\circ}_{\mathrm{D}}(t); t\right) \ = \ DC^{N}\left(C^{\circ}_{\mathrm{D}}(t); t\right) \ := \ DC^{*}\left( \frac{C^{\circ}_{\mathrm{D}}(t)}{N(t)} \right) \text{,}
\end{equation}
or, alternatively,
\begin{equation}
    \label{eq:nonlineardiscounting:relativetoGDP}
    DC\left(C^{\circ}_{\mathrm{D}}(t); t\right) \ = \ DC^{Y}\left(C^{\circ}_{\mathrm{D}}(t); t\right) \ := \ DC^{*}\left( \frac{C^{\circ}_{\mathrm{D}}(t)}{Y(t)} \right) \text{.}
\end{equation}
The latter approach allows us to penalize damages that exceed a certain percentage of the GDP.

The factor $DC^{*}(x)$ depends on the size of $x$. For small $x$ we have $DC^{*}(x) = 1$, but for large $x$ we may have factors $> 1$. Obviously, this will penalize large spikes in the costs.

\section{Model Implementation}

We provide a complete open-source implementation \cite{finmath-climate} of the model illustrating all our extensions, utilizing the Monte-Carlo implementation of standard interest rate models for simulation and (stochastic) algorithmic differentiation to efficiently calculate the sensitivities of model quantities against each other.

Our implementation can deal with an arbitrary time-discretization $\{ t_{i} \}$ and the DICE model is redefined as the Euler-discretization of a time-continuous model. For the equidistant discretisation with $\Delta t_{i} = 5~\text{years}$ the classical DICE-2016 model is recovered.  Our numerical experiments are based on a time-discretization with $\Delta t_{i} = 1~\text{year}$, however, the differences to the $\Delta t_{i} = 5~\text{years}$ discretization of the classical DICE-2016 are negligible in our study.
With annual time-discretization the piecewise constant abatement policy constitutes $499$ free parameters for a $T = 500~\text{years}$ model.

In the numerical experiments an annual time-discretization $\Delta t_{i} = 1$ was

%TODO: otpional: do also robustness simulations with delta_T = 5

%%%%%%%%%%%%%%%%%%%%%%%%%%%%%%%%%%%%%%%%%%%%%%%%%%%%%%%%%%%%%%%%%%%%%%%%%%%%%%%%%%%%%%%%%%%%%%%%%%%

\clearpage
\section{Associated Concepts}

\subsection{Time Preference Rates and the Ramsey Equation}
\label{sec:ramseyequation}

The DICE model extends a standard neoclassical optimal growth model (the \textit{Ramsey model}) with economic impacts from climate change.

We give a short recapitulation of the Ramsey equation since it makes clear that our modelling of a stochastic \textit{time preference rate} $\rho$ is equivalent to a stochastic {social discount rate} $r$.

\bigskip

Given that the utility $U$ is a function of the time-$t$ consumption $C(t)$, the discounted marginal utility is
\begin{equation*}
    U^{\prime}(C(t)) \cdot \exp(- \rho t) \text{.}
\end{equation*}
If we assume a declining marginal utility (constant relative risk aversion)
\begin{equation*}
    \mathrm{d} U^{\prime} / U^{\prime} \ = \ -\eta \ \mathrm{d} C / C \text{,}
\end{equation*}
where $\eta$ denotes the elasticity of marginal utility, then we have
\begin{align*}
    & \frac{\mathrm{d}}{\mathrm{d} t} \left( U^{\prime}(C(t)) \cdot \exp(- \rho t) \right) \\
    & \ = \ -\eta \cdot U^{\prime}(C(t)) \cdot \frac{\mathrm{d} C(t)}{\mathrm{d}t} / C(t) \cdot \exp(- \rho t) - \rho \cdot U^{\prime}(C(t)) \cdot \exp(- \rho t) \\
    & \ = \ - (\eta \cdot g + \rho) \cdot U^{\prime}(C(t)) \cdot \exp(- \rho t)
    \ = \ - r \cdot U^{\prime}(C(t)) \cdot \exp(- \rho t) \text{.}
\end{align*}
with
\begin{equation}
    \label{eq:ramsey:ramsey_equation}
    r \ = \ \eta \cdot g + \rho \text{,}
\end{equation}
where $g = \frac{\mathrm{d} C(t)}{\mathrm{d}t} / C(t) = \frac{\mathrm{d}}{\mathrm{d}t} \ln(C(t))$.
In this context, $r$ is called the social discount rate and Equation~\eqref{eq:ramsey:ramsey_equation} is the \textit{Ramsey} equation.

The (optimal) growth of the consumption $g$ depends on the optimal emission and savings paths, and is an outcome of the model optimization.

The elasticity of marginal utility $\eta$ is an input to the model and usually empirically estimated \cite{buchholz_discounting_2010}.
For the link between \textit{time preference} and \textit{social discount rate} see also \cite{Hansel2020}.
The time-preference $\rho$ is an exogenous parameter and subject to controversial discussions.

That $\rho$ is perceived as exogenous is already transparent from Nordhaus 2017 paper on the DICE model \cite{Nordhaus2017}, showing a dependency of the social cost of carbon on the chosen (growth-adjusted) discount rate.

In \cite{newell_pizer_discounting_2004}
an analysis is performed on the dependency of the social cost of carbon on a distribution of the discount rate. Here, the Bond market interest rate is used as an input to the model.

Via the \eqref{eq:ramsey:ramsey_equation} we see with a constant given that $\eta$ a stochastic \textit{time preference rate} $\rho$ corresponds to a stochastic {social discount rate} $r$.

% \todo{Read this paper more carefully}

While~\cite{newell_pizer_discounting_2004} considers a distribution of discount rates derived from Bond market rates, there is no feedback effect between the integrated assessment model and the given interest rates.

The novelty of our extensions is to consider a stochastic process of future uncertain discount rates and consider its interaction with an integrated assessment model.
This models the possibility that policymakers adapt the abatement strategy to changes in interest rates.
Such an adaptation is plausible, because already the static analysis shows that the optimal abatement strategy depends on the level of the interest rate.

%%%%%%%%%%%%%%%%%%%%%%%%%%%%%%%%%%%%%%%%%%%%%%%%%%%%%%%%%%%%%%%%%%%%%%%%%%%%%%%%%%%%%%%%%%%%%%%%%%%

\subsection{Social Cost of Carbon}

A prominent example of a metric describing the temporal structure of cost is the \textit{social cost of carbon} (SCC).

Model extensions updating IAMs \cite{Glanemann2020, Hansel2020, dietz_economic_2021,taconet_social_2021} show a wide range of potential SCC, while also econometrics-based SCC estimates for single impacts such as mortality \cite{carleton_valuing_2020} or energy consumption \cite{rode_estimating_2021}, but also general SCC \cite{rennert_comprehensive_2022} emphasize the need to improve the precision of the estimated SCC, especially by including additional impact channels. 

The social cost of carbon is defined as
\begin{equation}
	\label{eq:scc_definition}
    SCC(t) \ := \ -1000 \cdot \frac{\partial V(t)}{\partial E(t)} \Big/ \frac{\partial V(t)}{\partial C(t)} \text{.}
\end{equation}
The $SCC$ is the time-$t$ \textit{marginal cost} of emitting one additional unit of carbon. Its unit is $[SCC] = \frac{\mathrm{USD}}{\mathrm{tCO}_{2}}$.
We would like to note that depicting $SCC$ in the form~\eqref{eq:scc_definition} may be misleading, since it is an undiscounted time-$t$ value. Instead, one should rather consider the discounted, or, numéraire relative $SCC^{N}$, i.e.,
\begin{equation*}
    SCC^{N}(t) \ := \ -\frac{1000}{N(t)} \cdot \frac{\partial V(t)}{\partial E(t)} \Big/ \frac{\partial V(t)}{\partial C(t)} \text{.}
\end{equation*}

The ``social cost of carbon'' is maybe not a suitable measure to access the temporal distribution of the social burden, and hence intergenerational equity, because pricing emissions at the level of the SCC does not cover the cost associated with the climate mitigation paths: the accumulated value $\int_{0}^{T} SCC(t) \cdot E(t) \cdot \frac{N(0)}{N(t)} \mathrm{d}t$ does not match the accumulated cost $\int_{0}^{T} C(t) \cdot \frac{N(0)}{N(t)} \mathrm{d}t$, see \cite{Fries2023CO2}.

%TC:endignore

\end{document}